\documentclass[letterpaper,twocolumn,aps,prl,amsmath,groupedaddress,floatfix]{revtex4-1}
\usepackage{amsmath}
\usepackage{amssymb}
\usepackage{mathdots}
\usepackage{graphicx}
\usepackage{hyperref}

\makeatletter

\pdfpageheight\paperheight
\pdfpagewidth\paperwidth

\usepackage{color}\usepackage{verbatim}
\usepackage{amsfonts}
\usepackage{epsfig}
\usepackage{braket}
\usepackage{bm}
\graphicspath{{Figures/}}

\makeatother

\begin{document}
\title{Flat Band Josephson Junctions with Quantum Metric}
\author{Zhong C.F. Li$^{1}$}\thanks{These authors contributed equally to this work.}
\author{Yuxuan Deng$^{1}$} \thanks{These authors contributed equally to this work.}
\author{Shuai A. Chen$^{1}$} \thanks{chsh@ust.hk}
\author{Dmitri K. Efetov$^{2}$} 
\author{K. T. Law$^{1}$}\thanks{phlaw@ust.hk}
\affiliation{1. Department of Physics, Hong Kong University of Science and Technology, Clear Water Bay, Hong Kong, China}
\affiliation{2. Fakult\"at f\"ur Physik, Ludwig-Maximilians-Universit\"at, Schellingstrasse 4, M\"unchen 80799, Germany}

\begin{abstract}  In this work, we consider superconductor/flat band material/superconductor (S/FB/S) Josephson junctions (JJs) where the flat band material possesses isolated flat bands with exactly zero Fermi velocity. Contrary to conventional S/N/S JJs where the critical Josephson current vanishes when the Fermi velocity goes to zero, we show in this work that the critical current in the S/FB/S junction is controlled by the quantum metric length $\xi_\mathrm{QM}$ of the flat bands. Microscopically, when $\xi_\mathrm{QM}$ of the flat band is long enough, the interface bound states originally localized at the two S/FB, FB/S interfaces can penetrate deeply into the flat band material and hybridize to form Andreev bound states (ABSs). These ABSs are able to carry long range and sizable supercurrents. Importantly, $\xi_\mathrm{QM}$ also controls how far the proximity effect can penetrate into the flat band material. This stands in sharp contrast to the de Gennes' theory for S/N junctions which predicts that the proximity effect is expected to be zero when the Fermi velocity of the normal metal is zero. We further suggest that the S/FB/S junctions would give rise to a new type of resonant Josephson transistors which can carry sizable and highly gate-tunable supercurrent.
\end{abstract}

\maketitle
\emph{Introduction.}---The study of flat band superconductors had attracted much attention in recent years due to the discovery of superconducting moir\'e materials with flat bands~\cite{bistritzer2011moire,cao2018unconventional,2018Natur.556...80C,2019Natur.574..653L,2019Sci...363.1059Y,2019PhRvB..99s5455P,2019SciA....5.9770C,2019PhRvX...9c1049H,balents2020superconductivity,2020PhRvB.102t1112H,2020Natur.583..375S,2020NatMa..19.1265A,2020NatPh..16..926S,2021NatRM...6..201A,2021PhRvL.126b7002P,park2021tunable,2021PNAS..11806744V,2021COSSM..25j0952R,torma2022superconductivity,tian2023evidence,PhysRevLett.132.026002,2023arXiv231015558D}. One interesting property of flat band superconductors is that the superfluid weight is not zero but proportional to the quantum metric~\cite{provost1980riemannian,cheng2010quantum, provost1980riemannian,RevModPhys.84.1419} of the flat band despite the vanishing Fermi velocity~\cite{peotta2015superfluidity}. This finding inspired a large number of studies on the superfluid weight of flat band superconductors~\cite{PhysRevLett.117.045303,PhysRevB.94.245149,2017PhRvB.95b4515L,torma2018quantum,PhysRevB.98.134513,2019PhRvL.123w7002H,balents2020superconductivity,xie2020topology,PhysRevB.101.060505,PhysRevB.103.144519,verma2021optical,PhysRevB.106.014518,2022PhRvL.128h7002H,PhysRevB.106.184507,2022arXiv220900007H,2022PhRvB.106a4518H,2022PhRvL.128h7002H,PhysRevLett.131.240001,PhysRevB.107.214508}. More recently, it was realized that the quantum metric defines an important electronic length scale in flat band materials called the quantum metric length (QML) $\xi$ ~\cite{hu2023anomalous}. In particular, $\xi$ determines the superconducting coherence length, which is expected to be zero for flat band superconductors according to BCS theory~\cite{PhysRevLett.132.026002, hu2023anomalous}. Linking the superconducting coherence length with $\xi$~\cite{PhysRevLett.132.026002, hu2023anomalous} could explain the observed long superconducting coherence length in twisted bilayer graphene~\cite{tian2023evidence}, which deviated greatly from what the BCS theory predicted. 

As the superconducting coherence length generally controls the size of electronic objects in superconductors, we would expect $\xi$  to also govern the size of electronic objects such as Andreev bound states (ABSs), Yu-Shiba states in flat band materials. In this work, we verify this speculation and demonstrate how the quantum metric can control the ABS size at the weak link of the superconductor/flat band material/superconductor (S/FB/S) Josephson junctions (JJs). Furthermore, we show that the superconducting proximity effect can penetrate deeply into the flat band material and the decay length is controlled by the QML, which can be several orders of magnitude longer than the lattice length scale. Our results stand in sharp contrast to the de Gennes' theory of superconductor/normal metal (S/N) junctions which predicts that the superconducting proximity effect into a flat band should be zero~\cite{de2018superconductivity,tafuri2019fundamentals}.  

\begin{figure}[t]
\includegraphics[width=1\linewidth]{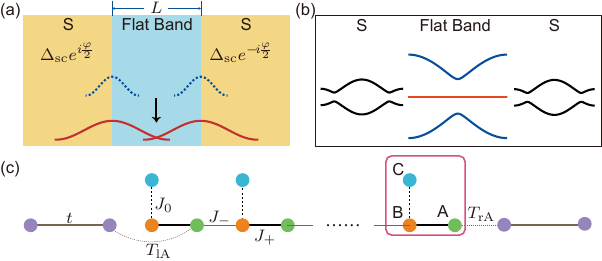} \caption{(a) A schematic illustration of a S/FB/S junction. The blue dashed lines denote two interface states in the small quantum metric regime.  When the quantum metric increases, the two interface states hybridize with each other to form ABSs at the junction as denoted by the solid red curves. (b) The schematic band structure of the flat band material and the superconducting leads. We assume an isolated flat band (red line) near the Fermi energy of the JJ. (c) The lattice model of the JJ. The flat band is described by a Lieb lattice with three lattice sites per unit cell. The superconducting leads couple to the nearest A-sites of the Lieb lattice with coupling strengths $T_{\mathrm{l} \mathrm{A}}$ and $T_{\mathrm{r} \mathrm{A}}$ respectively. The purple dots represent sites of the superconducting leads. }
\label{Fig1} 
\end{figure}


In the remaining sections of this work, we first use a one dimensional model to illustrate the properties of S/FB/S junctions (as schematically shown in Fig.~\ref{Fig1}(a) and (b)) and the conclusions can then be easily generalized to two dimensions. First, we build a model to describe the junction where the weak link is a 1D Lieb lattice~\cite{PhysRevLett.62.1201,PhysRevB.87.125428,PhysRevB.94.241409,slot2017experimental,PhysRevB.99.045107} which possesses a pair of spin degenerate flat bands and tunable quantum metric. The Lieb lattice model is illustrated in Fig.~\ref{Fig1}(c). We define the QML $\xi_\mathrm{QM}$ for the 1D Lieb model. Second, when the Lieb lattice is coupled to two superconductors and when the quantum metric is small, two interface states (denoted by blue dashed lines) are created at the S/FB and FB/S interfaces as shown in Fig.~\ref{Fig1}(a). Third, when the quantum metric increases, $\xi_\mathrm{QM}$ increases and the interface states penetrate deeper into the bulk of the Lieb lattice as the decay length of the interface state is controlled by $\xi_\mathrm{QM}$. When $\xi_\mathrm{QM}$ is comparable to the junction length, the two interface states hydridize to form two ABSs (illustrated by the red solid lines in Fig.~\ref{Fig1}(a)). As a result, the energy levels of these ABSs are sensitive to the phases of the superconductors so that they can carry supercurrents. Finally, we show that the critical Josephson current is sizable and highly gate-tunable even for long junctions and such S/FB/S JJs are new types of resonant Josephson transistors~\cite{beenakker1992resonant,kuhn2001supercurrents,RevModPhys.76.411,PhysRevB.95.014522,wen2019josephson}. 


\emph{QML $\xi_\mathrm{QM}$ of Lieb lattice---} To start with, we focus on the Lieb lattice which describes the weak link of the flat band JJ. The Lieb lattice model possesses two isolated, spin degenerate, flat bands near the Fermi energy as depicted in Fig.~\ref{Fig1}(b) and ~\ref{Fig2}(a). The Lieb lattice model has three sites per unit cell, labeled as A, B and C sites respectively [see Fig.~\ref{Fig1}(c)]. The Hamiltonian can be written as:
\begin{equation}
\begin{aligned}
H_{\mathrm{Lieb}}= & \sum_{i\sigma}( J_{+} a_{i\mathrm{A}\sigma}^{\dagger}a^{}_{i\mathrm{B}\sigma} + J_{0} a_{i\mathrm{C}\sigma}^{\dagger} a^{}_{i\mathrm{B}\sigma}  \\
 &  +  J_{-} a_{i-1\mathrm{A}\sigma}^{\dagger}a^{}_{i\mathrm{B}\sigma} + \mathrm{h.c.}) -\sum_{\alpha i \sigma}\mu_\mathrm{N}a_{i\alpha\sigma}^{\dagger}a^{}_{i\alpha\sigma}.\label{eq:Lieb}
\end{aligned}
\end{equation}
Here, $J_{\pm}=J(1\pm\delta)$, $J_0=\delta J$, $\alpha$ is the orbital index, $\mu_{\mathrm{N}}$ is the chemical potential. The flat band of the Lieb lattice in Eq.~(\ref{eq:Lieb}) is separated from the other two dispersive bands by an energy gap $\sqrt{2}J\delta$. The lattice constant is $a$ which is set to be unity. 


\begin{figure}[t]
\includegraphics[width=1\linewidth]{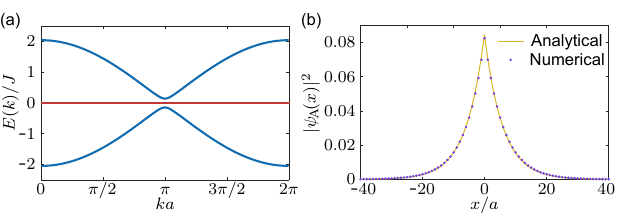} \caption{(a) The energy spectrum of a 1D Lieb lattice. The flat band (in red) is separated from the dispersive bands by $\delta J$. The parameters of the model are: $J=10^4\Delta_{\mathrm{sc}}$, $\delta=0.03$. The superconducting gap  $\Delta_{\mathrm{sc}}\!=\!1$ is set to be the energy unit for later discussions. (b) The probability distribution of the A-component of the bound state wavefunction $|\psi_\mathrm{A}(x)|^{2}$ localized near $x=0$ for a Lieb lattice with a local potential perturbation at $x=0$. The decay length of this bound state is well-fitted by the analytical expression of $\psi_\mathrm{A}(x) \propto \exp{(- |x|/8\xi_\mathrm{QM})}$. }
\label{Fig2} 
\end{figure}


It is important to note that there is a lack of natural electronic length scale for the flat bands as the length scales associated with $k_\mathrm{F}$ and $v_\mathrm{F}$ are either not well-defined or equal to zero. In the following section, we show how the QML of the 1D Lieb lattice $\xi_\mathrm{QM}$ governs the decay length of bound states near the flat band energy of the Lieb lattice. To find $\xi_\mathrm{QM}$, we start with the Bloch Hamiltonian of the Lieb lattice $h(k)$ (with the spin index omitted) and the Bloch states of the flat band $ \ket{u_0 (k)} $ which can be written as:
\begin{equation}\label{Lieb Bloch Ham}
   \!\!\!\! h(k)=2J
    \begin{pmatrix}
        0 & a_{k} & 0\\
        a_{k}^{*} & 0 & b_{k}\\
        0 & b_{k}^{*} & 0\\
    \end{pmatrix}, 
        \ket{u_0 (k)} =
    \begin{pmatrix}
        \frac{b_{k}}{\sqrt{|a_{k}|^2 + |b_{k}|^2}} \\
        0 \\
        -\frac{a_{k}^*}{\sqrt{|a_{k}|^2 + |b_{k}|^2}} \\
    \end{pmatrix}, 
\end{equation}
respectively. Here, $ a_{k} = \cos(\frac{ka}{2}) + i\delta\sin(\frac{ka}{2})$ and $b_{k} = i\delta$. For the 1D Lieb lattice, the real part of the quantum metric tensor associated with the flat band~\cite{provost1980riemannian} can be simplified as $g^{0}(k)$ such that 
\begin{equation}\label{quantum_metric}
    g^{0}(k)=\mathrm{Re}\langle\partial_{k}u_{0}(k)|(1-\vert u_{0}(k)\rangle\langle u_{0}(k)\vert)|\partial_{k}u_{0}(k)\rangle.
\end{equation}

With $g^{0}(k)$, the QML of the Lieb lattice is:
\begin{equation}\label{QML}
\xi_\mathrm{QM} = \int_{-\pi/a}^{\pi/a} g^{0}(k)\frac{dk}{2\pi}.
\end{equation}

For the Lieb lattice, and for small  $\delta$, we have $\xi_\mathrm{QM} = \frac{a}{16 \sqrt{2}\delta}$ where $\delta$ tunes the coupling strength between the C-sites and the B-sites in the Lieb lattice. It is important to note that the Lieb lattice only involves nearest neighbor hopping but $\xi_\mathrm{QM}$ can be much larger than $a$. For example, with $\delta =0.01$, $\xi_\mathrm{QM}$ is about $4.4a$. Next, we show that  $\xi_\mathrm{QM}$ dictates the size of the bound states which have energy near the flat band energy. For example, we introduce a single impurity at position $x=0$ at site A of the Lieb lattice to trap a bound state. It is shown in the Supplemental Materials (SM)~\cite{supp} that the bound state wavefunction $\psi_\mathrm{A}(x)$, which is localized near $x=0$, can be written as:
\begin{equation}\label{psiA}
    \begin{aligned}
        \psi_\mathrm{A}(x) \approx &\sum_{k} e^{ikx}|u_\mathrm{A}(k)|^2  \propto \exp{(- |x|/8\xi_\mathrm{QM})}.
    \end{aligned}
\end{equation}
Here, $u_\mathrm{A}(k) =  b_{k}/\sqrt{|a_{k}|^2 + |b_{k}|^2} $ is the first (A-site) component of $ \ket{u_0 (k)} $ and the decay length $8\xi_\mathrm{QM}$ is determined by the complex poles of $|u_\mathrm{A}(k)|^2$. Since the decay length generally depends on the wavefunctions of the flat band, we expect that the factor of $8$ is model dependent. A bound state wavefunction calculated numerically is shown in Fig.~\ref{Fig2}(b). Remarkably, the analytical solution in Eq.~\eqref{psiA}, with $\xi_\mathrm{QM}=\frac{a}{16 \sqrt{2}\delta}$, matches the numerical results of Fig.~\ref{Fig2}(b) extremely well and it clearly demonstrates that the size of the bound state wavefunction is controlled by $\xi_\mathrm{QM}$. The more general bound state solutions of the Lieb lattice are presented in the SM~\cite{supp}. Importantly, as shown below, the size of the ABSs of a flat band JJ are also governed by $\xi_\mathrm{QM}$ which determines the JJ properties.

\emph{Interface states of flat-band JJs.---} 
In this section, we study a flat band JJ as depicted in Fig.~\ref{Fig1}(a). The left (L) and right (R) leads are set to be conventional s-wave superconductors described by the Hamiltonians $H_{\mathrm{L}}$ and $H_{\mathrm{R}}$, respectively, where
\begin{align}
\!\!\!\!H_{\mathrm{L/R}}= & \!\!\sum_{\langle ij\rangle\sigma}-(t+\mu_\mathrm{s}\delta_{ij})c_{i\sigma}^{\dagger}c^{}_{j\sigma}+\!\sum_{i}(\Delta_{\mathrm{l/r}}c_{i\uparrow}^{\dagger}c_{i\downarrow}^{\dagger}+\mathrm{h.c.}). \label{eq:hsc}
\end{align}
Here, $\langle ij\rangle$ denotes the hopping between the nearest neighbor sites, $\sigma=\uparrow\downarrow$ is the spin index and $\mu_\mathrm{s}$ is the chemical potential.  The chemical potential is set to be $\mu_\mathrm{s} = \mu_\mathrm{N} =0 $ in this section so that the flat band energy is tuned to the Fermi energy of the superconducting leads. The pairing potentials of the two leads are denoted by $\Delta_{\mathrm{l/r}}=\Delta_{\mathrm{sc}}e^{\pm i\varphi/2}$ where $\varphi$ denotes the phase difference between the two superconductors and $\Delta_{\mathrm{sc}}$ is a constant. 

The couplings between the superconducting leads and the Lieb lattice (modelled by Eq.~\eqref{eq:Lieb}) is described by the coupling Hamiltonian $H_{c}$
\begin{equation}
H_{c}=\sum_{\alpha\sigma}(T_{\mathrm{l}\alpha}c_{\mathrm{l}\sigma}^{\dagger}a^{}_{\mathrm{r}\alpha\sigma}+T_{\mathrm{r}\alpha}c_{\mathrm{r}\sigma}^{\dagger}a^{}_{\mathrm{l}\alpha\sigma}+ \mathrm{h.c.}),\label{eq:contact_ham}
\end{equation}
where $T_{\mathrm{l/r},\alpha}$ labels the coupling between the left/right superconducting lead with the nearest $\alpha$-site of the Lieb lattice, and $\alpha=$A is chosen as an example. (See  Fig.~\ref{Fig1}(c)). Coupling to other sites will only change the results quantitatively. 

\begin{figure}[t]
    \includegraphics[width=1\linewidth]{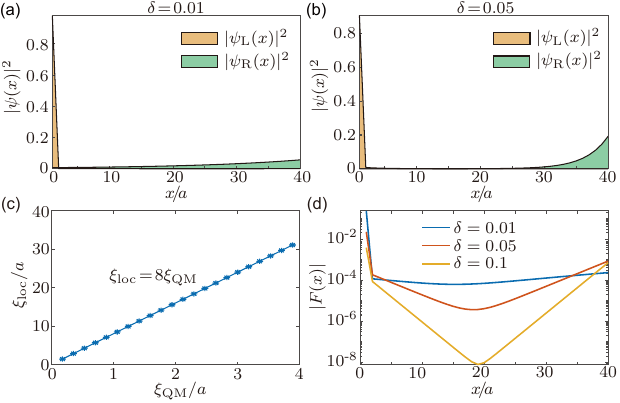} \caption{Probability distribution of the interface states $\Psi_\mathrm{L}(x)$ and $\Psi_\mathrm{R}(x)$ (a) $\delta=0.01$ and (b) $\delta=0.05$. The wavefunctions become more localized for larger $\delta$. The junction length is set to be $L=40$. The parameters of the Hamiltonian $H_\mathrm{JJ}$ are: $J\!=\!10^4\Delta_{\mathrm{sc}}$, $t\!=\!100\Delta_{\mathrm{sc}}$. (c) The localization length $\xi_\mathrm{loc}$ of the interface states for different values of $\xi_\mathrm{QM}$. The relation $\xi_\mathrm{loc}\!=\! 8 \xi_\mathrm{QM}$ was found both numerically and analytically. (d) The pairing correlation $|F(x)| \!=\!|\sum_{\alpha} \langle a_{x\alpha\uparrow}^{\dagger} a_{x\alpha\downarrow}^{\dagger} \rangle|$ in log scale. $|F(x)|$ decays exponentially into the bulk for large $\delta$. When $\delta$ is small, which corresponds to a large $\xi_\mathrm{QM}$, the pairing correlation can penetrate deeply into the bulk of the weak link.}
    \label{Fig3} 
\end{figure}


The total Hamiltonian of the one-dimensional flat band JJ can be written as $H_{\mathrm{JJ}}=H_{\mathrm{R}}+H_{\mathrm{L}}+H_{\mathrm{Lieb}}+H_{\mathrm{c}}$. In conventional S/N/S JJs, the critical supercurrent as well as the superconducting proximity effect are expected to be zero when the Fermi velocity of the normal metal goes to zero~\cite{klapwijk2004proximity,tafuri2019fundamentals}. Next, we demonstrate how an ABS which spreads across the JJ can emerge when the QML $\xi_\mathrm{QM}$ of the Lieb lattice is comparable to the junction length and these ABSs can carry sizable and long range supercurrents. The pairing correlation can also penetrate deeply into the junction.

As depicted in Fig.~\ref{Fig3}(a), when the two superconducting leads are coupled to the Lieb lattice with flat bands, two interface states, $\Psi_\mathrm{L}(x)$ and $\Psi_\mathrm{R}(x)$ are created as schematically shown in Fig.~\ref{Fig1}(b). The states $\Psi_\mathrm{L}(x)$ and $\Psi_\mathrm{R}(x)$ are not degenerate with each other in general, but each of them is spin degenerate. These wavefunctions have six components due to the electron and hole contributions from the three orbitals (A, B and C orbitals) of the Lieb lattice. Physically, as the middle band of the Lieb lattice is exactly flat and the superconductor is gapped, when an interface state with energy away from the flat band energy is created, this state must be localized at the interface. Given that the energy of the state is also within the quasiparticle gap of the superconducting leads, the question is, what is the localization lengths of these interface states? According to the de Gennes' theory of proximity effect~\cite{de2018superconductivity,tafuri2019fundamentals}, the localization length of the bound states should be zero because of the zero Fermi velocity of the flat bands.

To answer this question, we recall that the QML determines the superconducting coherence length of flat band superconductors~\cite{PhysRevLett.132.026002,hu2023anomalous}. It is reasonable to speculate that $\xi_\mathrm{QM}$  should be related to the localization length of the bound states and this is indeed the case. Fig.~\ref{Fig3}(a)-(b) depict $|\Psi_\mathrm{L}(x)|^{2}$ and $|\Psi_\mathrm{R}(x)|^2$ at the flat band JJ with two different values of $\delta$, respectively. Even though the two interface states look very different from each other as the exact forms of the wavefunctions depend on the details of the interfaces, the localization lengths of the wavefunctions (on the Lieb lattice side) are the same and are controlled by $\xi_\mathrm{QM}$. Fig.~\ref{Fig3}(c) depicts the localization lengths $\xi_\mathrm{loc}$ of the two interface states as a function of $\xi_\mathrm{QM}$. The localization length of the interface states is extracted by assuming that the wavefunction inside the weak link has a form $\Psi_\mathrm{L/R}(x) \propto e^{-|x|/\xi_\mathrm{loc}}$, where $x$ is measured from the S/FB or FB/S interfaces. It is clear from Fig.~\ref{Fig3}(c) that the numerically extracted localization length of the interface states is $\xi_\mathrm{loc} = 8\xi_\mathrm{QM}$ where $\xi_\mathrm{QM} = a/(16\sqrt{2}\delta)$. This is the same decay length found in Eq.~\eqref{psiA}, and analytical results are given in the SM~\cite{supp}. 

Importantly, when the $\xi_\mathrm{QM}$ is comparable to the junction length, the two interface states hybridize into two ABSs which spread across the weak link region. In this case, there is pairing correlation across the whole weak link. The pairing correlation is defined as $|F(x)| = |\sum_{\alpha} \langle a_{x\alpha\uparrow}^{\dagger} a_{x\alpha\downarrow}^{\dagger} \rangle|$, where $\alpha$ is the orbital index. $F(x)$ inside the flat band material can be easily extracted from the Green's function $G(E) = 1/(E-H_\mathrm{JJ})$ in real space and the results are shown in Fig.~\ref{Fig3}(d). $F(x)$ with three different values of $\delta$ are demonstrated.  It is clear that when $\xi_\mathrm{QM}$  is large, the superconducting proximity effect can penetrate deeply into the bulk of the flat band material. 

\begin{figure}[t]
    \includegraphics[width=1\linewidth]{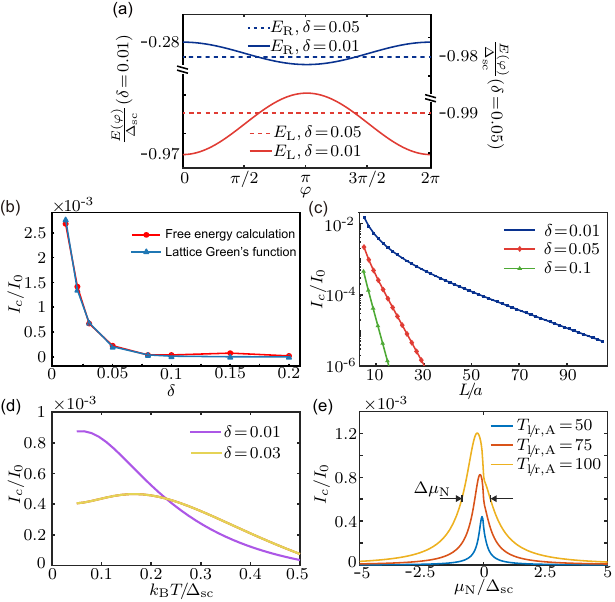} \caption{(a) The energy-phase relation of the ABS at the junction for $\delta=0.01$ and $\delta=0.05$, respectively. As $\delta$ increases, the bound state energy becomes insensitive to the change of $\varphi$. (b) The critical current as a function of $\delta$ with $L=10a$. The results obtained from the free energy calculations and lattice Green's functions agree with each other. (c) The critical current generally decays exponentially as the junction length $L$ increases. (a)-(c) are calculated with temperature $k_\mathrm{B}T\!=\!0.1\Delta_{\mathrm{sc}}$. (d) The critical current $I_c$ as a function of temperature $k_{\mathrm{B}}T$, for different choices of $\delta$. The above figures are plotted using $T_{\mathrm{r}/\mathrm{l},\mathrm{A}}\!=\!10^2\Delta_{\mathrm{sc}}$ and $J\!=\!10^4\Delta_{\mathrm{sc}}$. (e) The critical current $I_c(\mu_\mathrm{N})$ as a function of the chemical potential of the flat band, for different coupling strengths $T_{\mathrm{r}/\mathrm{l},\mathrm{A}}$. The temperature is set at $k_\mathrm{B}T=0.01\Delta_{\mathrm{sc}}$. The full width at half maximum, $\Delta \mu_\mathrm{N}$, is proportional to $|T_{\mathrm{r}/\mathrm{l},\mathrm{A}}|^2$. Figure(a), (d) and (e) are calculated with $L=20a$. The unit of current is $I_0=\frac{e\Delta{\mathrm{sc}}}{\hbar}$.}
    \label{Fig4} 
\end{figure}

\emph{ABSs and Josephson Currents---} As shown in Fig.~\ref{Fig3}(c) and schematically illustrated in Fig.~\ref{Fig1}(a), when $\xi_\mathrm{Lieb}=8\xi_{QM}$ is comparable to the junction length, the two interface states hybridize into two ABSs which spread across the weak link. As a result, the energy levels of the ABSs would depend on the phase difference of the two superconducting leads $\varphi$. Therefore, we would expect a finite Josephson current  $I= \frac{\partial F_\mathrm{JJ}(\varphi)}{\partial \varphi}$ where $F_\mathrm{JJ}(\varphi)$ is the free energy of the JJ. In Fig.~\ref{Fig4}(a), the energy of the ABSs as a function of $\varphi$ are shown. When $\xi_\mathrm{QM}$ is short as compared to the junction length (large $\delta$), the two interface states do not couple to each other and the bound state energies are insensitive to the change of $\varphi$ (dashed lines in Fig.~\ref{Fig4}(a)).  On the contrary, when $\xi_\mathrm{QM}$ is long (with small $\delta$), the two interface states hybridize. In this case, the energies of the two ABSs $E_{\pm}(\varphi)$ are $\varphi$ dependent (solid lines) and finite Josephson currents would emerge. The critical Josephson current at different values of the QML is depicted in Fig.~\ref{Fig4}(b). The Josephson currents in Fig.~\ref{Fig4}(b) are determined by numerically evaluating $F_\mathrm{JJ}(\varphi)$ through exact diagonalization of a finite system. The same results are also obtained by the lattice Green's function approach~\cite{Datta_1995,FURUSAKI1994214}:
\begin{equation}
\begin{aligned}
I(\varphi) = & \frac{2e}{\hbar}k_\mathrm{B}T\mathrm{Im}\sum_{\omega_m}\mathrm{Tr}(V_{n+1,n}G_{n,n+1}(i\omega_m)- \\ 
& V_{n,n+1}G_{n+1,n}(i\omega_m)), \\
\end{aligned}
\end{equation}
where $\omega_m$ is the Matsubara frequency, $n$ is the index of lattice site, $V$ encodes the hopping amplitudes between neighboring sites and $[G_{n,n+1}(i\omega_m)]_{\alpha\beta}=\langle n\alpha|G(i\omega_m)|n+1,\beta\rangle$ refers to the matrix element of the lattice Green's function $G(i\omega_m)
$ of the entire Josephson junction with $G(i\omega_m)=(i\omega_m-H_\mathrm{JJ})^{-1}$. The details can be found in the SM~\cite{supp}.

The length dependence of the critical currents at three different QML is shown in Fig.~\ref{Fig4}(c). It is important to note that for large $\xi_\mathrm{QM}$, the critical current can be in the order of $10^{-2}$ to $10^{-3}$ of $I_{0}=\frac{e \Delta_\mathrm{sc}}{\hbar}$, which is the maximum Josephson current of a single conducting channel~\cite{beenakker1992three}. Interestingly, in our case, the large critical Josephson current can appear even if the band is exactly flat. This is one of the key results of this work. Furthermore, the temperature dependence of the Josephson current is shown in Fig.~\ref{Fig4}(d). It is interesting to note that the Josephson current is nonmonotonic as a function of temperature. This unexpected behavior can be easily understood by the out-of-phase energy-phase relations of the two ABSs [See Fig~\ref{Fig4}(a)].

\emph{Resonant Josephson Transistor ---} The Josephson current calculated above assumes that the energy of the flat bands lies at the Fermi energy of the superconductors. In this section, we show that the critical Josephson current decreases dramatically once the flat band energy is gated away from the Fermi energy of the superconductors. The critical current as a function of the chemical potential of the Lieb lattice is determined by a Breit-Wigner transmission probability function~\cite{Beenakker_JJ} as shown in Fig.~\ref{Fig4}(e). It is clear from Fig.~\ref{Fig4}(e) that the critical Josephson current has a sharp resonance peak when the ABS energy matches the Fermi energy of the superconducting leads. Moreover, the full width at half maximum (FWHM) of the resonance peak is proportional to the amplitude of the coupling strength between the superconducting leads and the Lieb lattice such that $\Delta \mu_{\mathrm{N}} \propto |T_{\mathrm{r}/\mathrm{l},\mathrm{A}}|^2$. This sharp resonance can lead to a new design of resonant Josephson transistors~\cite{beenakker1992resonant,kuhn2001supercurrents,PhysRevB.95.014522,wen2019josephson} in which a long range Josephson current is highly gate-tunable.  Importantly, the theory discussed in this work can apply to two-dimensional Lieb lattice mediated Josephson effect as well. 

\emph{Discussion and Conclusion.}--- Before concluding, we would like to discuss a few points. First, the Lieb lattice is chosen to model the flat band material because the lattice is incredibly simple which only involves nearest neighbor hopping. At the same time, the model can give rise to a very long $\xi_\mathrm{QM}$ which far exceeds the lattice length scale. This can avoid the complications induced by long range hopping. Second, the conclusion that the quantum metric can tune the critical Josephson current in S/FB/S junctions is generally true. We demonstrated the finite Josephson current with another flat band model in the SM~\cite{supp}. Moreover, the Lieb lattice has an energy spectrum which closely resembles the case of twisted bilayer graphene (TBG) near magic angle~\cite{2019PhRvB..99s5455P}. Our S/FB/S junction results are highly relevant to S/TBG/S JJs which were fabricated experimentally~\cite{2021NatNa..16..769R,de2021gate,PhysRevLett.131.016003,2023NatCo..14.2396D,PhysRevResearch.5.023029, PhysRevResearch.5.023029, PhysRevLett.130.266003, sun2023anomalous, li2023valley}. We would like to point out that the flat band Josephson effect had been studied previously in Ref.~\cite{PhysRevB.103.144519} but the quantum metric effect was not investigated.

To conclude, we show that the quantum metric, through the QML $\xi_\mathrm{QM}$, plays an extremely important role in determining the properties of S/FB/S JJs.  In sharp contrast to conventional S/N/S junctions where the critical supercurrent is expected to be zero when the Fermi velocity of the normal metal goes to zero, there can be a large Josephson current in S/FB/S junctions which is tuned by the quantum metric. Also, we demonstrated that the superconducting proximity effect can penetrate deep into the weak link even when the Fermi velocity of the weak link is zero. This also stands in sharp contrast to the well-established theory of de Gennes on superconducting proximity effects~\cite{de2018superconductivity}. The understanding of S/FB/S JJs would lead to a new design of highly gate-tunable resonant Josephson transistors.

\emph{Acknowledgement}---  We thank Tai-Kai Ng, Adrian Po and A. D\' iez-Carl\'on for valuable discussions. K. T. L. acknowledges the support of the Ministry of Science and Technology, China and the Hong Kong Research Grants Council through Grants No. 2020YFA0309600, No. RFS2021-6S03, No. C6025-19G, No. AoE/P-701/20, No. 16310520, No. 16310219, No. 16307622, and No. 16309223.
D.K.E. acknowledges funding from the European Research Council (ERC) under the European Union's Horizon 2020 research and innovation program (grant agreement No. 852927), the funding from the EU EIC Pathfinder program, project FLATS (grant agreement No. 101099139) and the Keele Foundation.

\bibliographystyle{apsrev4-2}
\bibliography{ref}

\clearpage
\onecolumngrid

\begin{center}
    \textbf{\large Supplemental Material for ``Flat Band Josephson Junctions with Quantum Metric"}\\[.2cm]
    Zhong C.F. Li$^{1}$, Yuxuan Deng$^{1}$, Shuai A. Chen$^{1}$, Dmitri K. Efetov$^{2}$, K. T. Law$^{1}$\\[.1cm]
    {\itshape ${}^1$Department of Physics, Hong Kong University of Science and Technology, Clear Water Water Bay, 999077 Hong Kong, China}\\
    {\itshape ${}^2$Fakult\"at f\"ur Physik, Ludwig-Maximilians-Universit\"at, Schellingstrasse 4, M\"unchen 80799, Germany}\\[1cm]
\end{center}

\maketitle

\setcounter{equation}{0}
\setcounter{section}{0}
\setcounter{figure}{0}
\setcounter{table}{0}
\setcounter{page}{1}
\renewcommand{\theequation}{S\arabic{equation}}
\renewcommand{\thesection}{ \Roman{section}}

\renewcommand{\thefigure}{S\arabic{figure}}
\renewcommand{\thetable}{\arabic{table}}
\renewcommand{\tablename}{Supplementary Table}

\renewcommand{\bibnumfmt}[1]{[S#1]}
\renewcommand{\citenumfont}[1]{#1}
\makeatletter

\maketitle

\setcounter{equation}{0}
\setcounter{section}{0}
\setcounter{figure}{0}
\setcounter{table}{0}
\setcounter{page}{1}
\renewcommand{\theequation}{S\arabic{equation}}
\renewcommand{\thesection}{ \Roman{section}}

\renewcommand{\thefigure}{S\arabic{figure}}
\renewcommand{\thetable}{\arabic{table}}
\renewcommand{\tablename}{Supplementary Table}

\renewcommand{\bibnumfmt}[1]{[S#1]}
\makeatletter

\section{Quantum Metric of the 1D Lieb Lattice}

\begin{figure}[ht]
    \centering
    \includegraphics[scale=0.8]{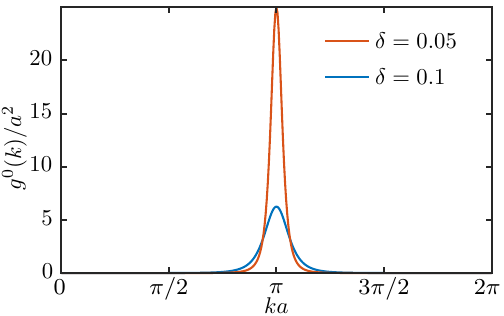}
    \caption{Distribution of quantum metric tensor $g^0(k)$ in the first Brillouin zone for different $\delta=0.05$ and $0.1$ for 1D Lieb lattice with Hamiltonian in Eq.~\eqref{Lieb_k}.}
    \label{QM1D}
\end{figure}

Here, we study the quantum metric of a 1D Lieb lattice. The Lieb lattice possesses three orbitals per unit cell and the Bloch Hamiltonian $H(k) =\sum_{\alpha\beta} a^\dagger_{\alpha k} h_{\alpha\beta}(k)a_{\beta k}$ has a flat band and the correponding Bloch state is denoted as $|u_0(k)\rangle$:
\begin{equation}\label{Lieb_k}
   \!\!\!\! h(k)=2J
    \begin{pmatrix}
        0 & a_{k} & 0\\
        a_{k}^{*} & 0 & b_{k}\\
        0 & b_{k}^{*} & 0\\
    \end{pmatrix}, \quad
    \ket{u_0 (k)} =
    \begin{pmatrix}
        \frac{i\delta}{\sqrt{2\delta^2+(1-\delta^2)\cos^2 \frac{ka}{2}}} \\
        0 \\
        \frac{-\cos(\frac{ka}{2}) + i\delta\sin(\frac{ka}{2})}{\sqrt{2\delta^2+(1-\delta^2)\cos^2 \frac{ka}{2}}} \\
    \end{pmatrix}, 
\end{equation}
with $ a_{k} = \cos(\frac{ka}{2}) + i\delta\sin(\frac{ka}{2})$ and $b_{k} = i\delta$. In 1D, the quantum geometric tensor is a scalar, and the quantum metric only has one component, 
\begin{align}
    g^{0}(k)=&\mathrm{Re}\langle\partial_{k}u_{0}(k)|(1-\vert u_{0}(k)\rangle\langle u_{0}(k)\vert)|\partial_{k}u_{0}(k)\rangle \notag \\
    =& \frac{\delta^2 a^2}{4} \frac{\sin^2 \frac{ka}{2} + \delta^2 \cos^2\frac{ka}{2}} {\left[2\delta^2+(1-\delta^2)\cos^2 \frac{ka}{2}\right]^2}.
\end{align}
For small $\delta$, the quantum metric tensor is peaked at $k=\pi/a$ with the maximum $\frac{a^2}{16\delta^2}$, as shown in Fig.~\ref{QM1D}.
The quantum metric length $\xi_{\mathrm{QM}}$ can be obtained by integrating over the first Brillouin zone, 
\begin{equation}\label{QM_length}
    \xi_{\mathrm{QM}} = \int_{-\pi/a}^{\pi/a} g^{0}(k)\frac{dk}{2\pi} = \frac{a}{16\sqrt{2}\delta} \frac{2\delta^4 + \delta^2 + 1}{(1+\delta^2)^{3/2}}.
\end{equation}
In the focus of the main text where $\delta$ is assumed to be small, the quantum metric length behaves $\xi_{\mathrm{QM}} = \frac{a}{16\sqrt{2}\delta}$, which becomes larger for a smaller $\delta$.

\section{Derivation of Flat Band Bound State Wavefunctions}\label{Analytical Model}
Below, we present the details of the bound states associated with the flat band in the Lieb lattice. We can apply an effective one-band description by means of a projection onto the flat band.
A bound state can be formed by introducing an on-site impurity at a single site of the lattice. Our target is to derive an analytical expression for the wavefunction of such a bound state. The following discussion applies to general flat-band systems.



We consider an infinite 1D system with an impurity on the original site $x=0$. The total Hamiltonian is $H=H_0+V$ with $V$ describing the on-site impurity. 
The $H_0$ possesses a flat band which is isolated from other dispersive bands. We use $|nk\rangle$ to label the eigenstates of $H_0$: $H_0= \sum_{nk} \epsilon_n(k)|nk\rangle\langle nk|$, where $n$ is the band index and $k$ is the wave vector. On the orbital basis $|\alpha k\rangle$, we have the components $u_{n\alpha}(k)$ defined by $|nk\rangle=\sum_{\alpha}u_{n\alpha}(k)|\alpha k\rangle$. 
In particular, we specify the wavefunction of a flat band by $|\bar{n}k\rangle$. 
For the on-site impurity, we consider the potential to be diagonal for orbitals,
\begin{align}
V= \sum_\alpha V_\alpha a^\dagger_{\alpha x=0} a_{\alpha x=0}.
\end{align}
The potential has matrix elements $\langle\alpha k|V|\beta k^\prime \rangle=\frac{V_\alpha}{N}\delta_{\alpha\beta}$ with $N$ being the number of sites. Moreover, we are interested in the case where the impurity strength is much weaker than the gap between the flat band and other ones. This assumption will allow us to ignore the dynamical effects of the other bands on the bound states.


The bound state can be resolved by the Schr\"odinger equation, 
\begin{align}
(H_0+V)|\psi\rangle=E|\psi\rangle.
\label{sm:scheq}
\end{align}
Initially, we project the Schr\"odinger equation in Eq.~\eqref{sm:scheq} onto the subspace spanned by the flat band states $\{|\Bar{n}k\rangle\}$:
\begin{align}\label{equation}
    \langle\Bar{n}k|(H_0+V)\sum_{k'}|\Bar{n}k'\rangle\langle\Bar{n}k'|\psi\rangle=&E\langle\Bar{n}k|\psi\rangle,\\
\sum_{k'}\langle\Bar{n}k|V|\Bar{n}k'\rangle\phi_\mathrm{k'}=&E\phi_{k}.
\end{align}
In the second line, we introduce the notation $\phi_{k}=\langle\Bar{n}k|\psi\rangle$. Therefore, we have reduced the multiband problem to an effective one-band problem. The next step is to solve the eigenvalue and eigenvector of the matrix $\mathbf{V}_{kk'}=\langle\Bar{n}k|V|\Bar{n}k'\rangle$, with 
\begin{equation}\label{V}
    \langle\Bar{n}k|V|\Bar{n}k'\rangle=\sum_{\alpha\beta}u^*_{\bar n\alpha}(k)u_{\bar n\beta}(k')\langle\alpha k|V|\beta k'\rangle=\frac{1}{N}\sum_{\alpha}V_{\alpha}u^*_{\bar n\alpha}(k)u_{\bar n\alpha}(k'),
\end{equation}
where $u_{\Bar{n}\alpha}$ is the Bloch wave of the flat band. Generally, the eigenvalue and eigenvector can be easily obtained numerically.
After that, the bound state wavefunction $|\psi\rangle$ and $\psi_{\alpha}(x)$, which is its real space component on orbital $\alpha$,  can be expressed as
\begin{align}
\label{psi}
|\psi\rangle&=\sum_{k}|\Bar{n}k\rangle\langle\Bar{n}k|\psi\rangle=\sum_{k}\phi_{k}|\Bar{n}k\rangle, \\
        \psi_{\alpha}(x)&=\frac{1}{\sqrt{N}}\sum_{k}e^{ikx}\phi_{k}u_{\bar n\alpha}(k).
 \end{align}
For an analytical form, we can reformulate Eq.~\eqref{equation} as
    \begin{align}
    \label{phik}
        E\phi_{k}=&\frac{1}{N}\sum_{k'}\sum_{\alpha}V_{\alpha}u^*_{\bar n\alpha}(k)u_{\bar n\alpha}(k')\phi_{k'} \notag\\
        =&\frac{1}{N}\sum_{\alpha}V_{\alpha}u^*_{\alpha}(k)\sum_{k'}u_{\bar n\alpha}(k')\phi_{k'} \notag \\
        =&\frac{1}{\sqrt{N}}\sum_{\alpha}V_{\alpha}u^*_{\bar n\alpha}(k)\psi_{\alpha}(x=0).
    \end{align}
We can construct the bound state wavefunction once we get the wavefunction components $\psi_\alpha(x=0)$ which satisfy the equation up to a normalization factor,
\begin{equation}
\psi_{\alpha}(x=0)= \frac{1}{NE}\sum_{k}\sum_{\beta}V_{\beta}u^*_{\bar n\beta}(k)\psi_{\beta}(x=0)u_{\bar n\alpha}(k).
\end{equation}
where the bound state energy $E$ and $\psi_\alpha(x=0)$ are the eigenvalue and the eigenvector of the matrix $U_{\alpha\beta} =\frac{1}{N}\sum_k V_\beta u_{\bar n \beta}(k)u_{\bar n \alpha }(k)$, respectively. 
And the total wavefunction is
\begin{align}
        \psi_{\alpha}(x)=&\frac{1}{NE}\sum_{k}e^{ikx}\sum_{\beta}V_{\beta}u^*_{\bar n\beta}(k)\psi_{\beta}(x=0)u_{\bar n\alpha}(k).
\end{align}

We are ready to apply this formula to solve flat band bound states in the 1D Lieb lattice. As discussed in the main text, we consider the potential at the site A of the original point. Then up to a normalization factor, we have 
\begin{align}
\label{psiA}
        \psi_{\mathrm{A}}(x)\propto&\frac{1}{NE}\sum_{k_x} e^{ik_xx}|u_{\mathrm{A}}(k_x)|^2\notag\\
        =&\frac{1}{NE}\sum_{k_x} e^{ik_x x}\frac{\delta^2}{\delta^2+\cos^2(k_x/2)+\delta^2\sin^2(k_x/2)} \notag\\
        \approx& \frac{1}{NE}\sum_{k_x}e^{ik_xx}\frac{\delta^2}{2\delta^2+(k_x-\pi)^2/4}\notag\\
        \propto&\frac{\delta}{\sqrt{2}}\exp{(-2\sqrt{2}\delta |x|+i\pi x)}.
\end{align}

\begin{figure}[t]
    \centering
    \includegraphics[scale=0.8]{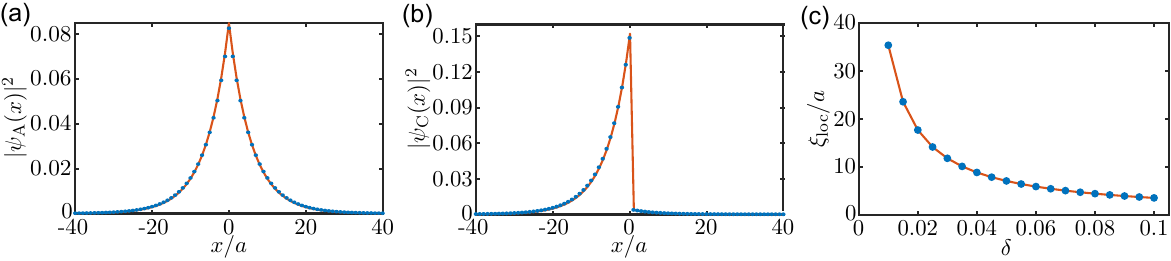}
    \caption{The distribution of wavefunctions of sublattices: (a) $\psi_{\mathrm{A}}$ and (b)  $\psi_{\mathrm{C}}$. (c) Localization length $\xi_{\mathrm{loc}}$ as a function of $\delta$.
    Here, the dots in blue are wavefunction components from numerically diagonalizing the Hamiltonian in Eq.~\eqref{sm:scheq} and the continuous lines are analytical results. Here the potential is acted on site of the original point $x=0$ only on the sublattice A. We conduct the numerical calculations with parameters $\delta=0.03$, $V=10$ and $J=1000$. 
    } 
    \label{BS_A,C}
\end{figure}

Here we evaluate the integral in the small $\delta$ limit. 
The localization length of $\psi_{\mathrm{A}}(x)$ is $1/(2\sqrt{2}\delta)$, so a smaller $\delta$ implies a longer localization length. In Fig.~\ref{BS_A,C}, the localization length and wavefunction predicted by the analytical model are compared with those obtained by diagonalizing the tight-binding Hamiltonian of the Lieb lattice.
Similarly, we can solve $\psi_{\mathrm{C}}$, and up to a normalization factor, we have 
\begin{align}\label{psiC}
    \psi_{\mathrm{C}}(x)\propto&\frac{1}{NE}\sum_{k_x} e^{ik_xx}u^*_{\mathrm{A}}(k)u_{\mathrm{C}}(k_x) \notag\\
        =&\frac{1}{NE}\sum_{k_x} e^{ik_xx}\frac{i\delta\cos(k_x/2)+\delta^2\sin(k_x/2)}{\delta^2+\cos^2(k_x/2)+\delta^2\sin^2(k_x/2)}\notag \\
       \propto &  \Theta(x)\frac{\sqrt{2}-1}{\sqrt{2}+1}\exp{\left(-2\sqrt{2}\delta x+i\pi x\right)}+\Theta(-x)\exp{\left(2\sqrt{2}\delta x+i\pi x\right)},
\end{align}
Hence, $\psi_{\mathrm{C}}(x)$ has discontinuity at $x=0$. The localization length of $\psi_{\mathrm{C}}$ is the same as $\psi_{\mathrm{A}}$. The comparison between the wavefunction and the localization length of $\psi_{\mathrm{C}}$ obtained from the analytic model and tight-binding model is shown in Fig.~\ref{BS_A,C}.

\begin{figure}[t]
    \centering
    \includegraphics[scale=0.8]{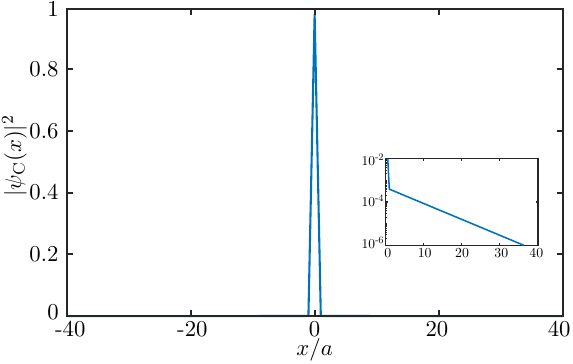}
    \caption{The magnitude of the wavefunction $\psi_{\mathrm{C}}$ at $\delta=0.03$ when $V$ is added on sublattice C, while parameter $J$ needs to be large enough to ensure that the energy scale of potential $V$ is much smaller than the band gap. The inset figure shows an exponential tail of the wavefunction away from the on-site potential. We conduct the numerical calculations with parameters $\delta=0.03$, $V=10$ and $J=1000$. 
    }
    \label{wf_C_impurity}
\end{figure}

If $V$ is added on the C site, meaning that only $V_{\mathrm{C}}$ is nonzero, then
    \begin{align}
        \psi_{{\mathrm{C}}}(x)\propto&\frac{1}{NE}\sum_{k_x} e^{ik_xx}|u_{\mathrm{C}}(k_x)|^2\notag\\
        =&\frac{1}{NE}\sum_{k_x} e^{ik_xx}\frac{\cos^2(k_x/2)+\delta^2\sin^2(k_x/2)}{\delta^2+\cos^2(k_x/2)+\delta^2\sin^2(k_x/2)}\notag\\
        =&\frac{1}{NE}\sum_{k_x} e^{ik_xx}\bigg(1-\frac{\delta^2}{\delta^2+\cos^2(k_x/2)+\delta^2\sin^2(k_x/2)}\bigg)\notag\\
    \propto&\delta(x)-\frac{\delta}{\sqrt{2}}\exp{(-2\sqrt{2}\delta |x|+i\pi x)}.
    \end{align}
where $\delta(x)$ is delta function.
This analytical expression aligns with numerical calculations [see Fig.~\ref{wf_C_impurity}]. When the impurity is only added at C site, the wavefunction is concentrated dominantly on one isolated C site. The magnitude of the tail, which also decays with a length $1/2\sqrt{2}\delta$, is extremely small.  Accordingly, the Josephson effect will be too weak to be manifested.


\section{Phenomenological Model for Flat Band Josephson Junctions}

In this section, we can use the Lieb lattice to discuss the Josephson current in the flat band Josephson junction. The weak link part of the Josephson junction is described by Hamiltonian $H_0$, which includes a flat band in the bulk, while the flat band energy is well separated from other dispersive bands. To emphasize the effect of the flat band, the chemical potential is chosen to be around the energy of the flat band rather than the dispersive band. By integrating out the superconducting part, we get an effective Hamiltonian in the weak link, which includes bulk Hamiltonian $H_0$ for $0<x<L$ and two self energy corrections $V_{\mathrm{L}}$ and $V_{\mathrm{R}}$ added on the leftmost site at $x=0$ and the rightmost site at $x=L$, respectively. The total BdG Hamiltonian of the junction can be expressed as
\begin{equation}
    H = H_0 + V_{\mathrm{L}} + V_{\mathrm{R}}.
    \label{eq: JJ Ham}
\end{equation}
For an isolated flat band in a multi-band system, the existence of a small local perturbation term would give rise to states with energy mismatch compared to the flat band state. In principle, $V_{\mathrm{L}}$ and $V_{\mathrm{R}}$ would give bound states localized at $x=0,L$, respectively. We can check from numerical calculations that for a positive half-infinite chain with the self energy correction at $x=0$, the wavefunction shares the same energy, decay length, and relative ratio between different components of the wavefunction on the $x$ positive half axis, and so does a negative half-infinite chain with self energy correction at $x=L$. Therefore, we can use states obtained in Sec.~\ref{Analytical Model} as trial wavefunctions to build a phenomenological model for the Josephson supercurrent.

In general, the self energy correction $V_{{\mathrm{L/R}}}$ can have both electron and hole components. This can be generalized straightforwardly by replacing $V_{\alpha}$ and $\phi_k$ in Eq.~\eqref{V} with 
\begin{equation}
    V_{\alpha} \rightarrow V_{\alpha}=
    \begin{pmatrix}
    V_{\alpha}^{\mathrm{ee}} & V_{\alpha}^{\mathrm{eh}} e^{i\varphi/2} \\
    V_{\alpha}^{\mathrm{he}} e^{-i\varphi/2} & V_{\alpha}^{\mathrm{hh}} \\
    \end{pmatrix},
    \quad \phi_k \rightarrow 
    \begin{pmatrix}
    \phi_k^{\mathrm{e}} \\
    \phi_k^{\mathrm{h}} \\
    \end{pmatrix}.
    \label{eq: substitution}
\end{equation}
The corresponding Schr\"{o}dinger equation becomes
\begin{equation}
    E
    \begin{pmatrix}
    \psi_{\alpha}^{({\mathrm{e}})}(x) \\
    \psi_{\alpha}^{({\mathrm{h}})}(x) \\
    \end{pmatrix}
    =\frac{1}{N} \sum_{k,\beta} e^{ikx}u_{\alpha}(k) u_{\beta}^*(k)
    \begin{pmatrix}
    V_{\beta}^{\mathrm{ee}} & V_{\beta}^{\mathrm{eh}} e^{i\varphi/2}\\
    V_{\beta}^{\mathrm{he}} e^{-i\varphi/2} & V_{\beta}^{\mathrm{hh}} \\
    \end{pmatrix}
    \begin{pmatrix}
    \psi_{\beta}^{({\mathrm{e}})}(x=0) \\
    \psi_{\beta}^{({\mathrm{h}})}(x=0) \\
    \end{pmatrix}.
    \label{eq: SE1}
\end{equation}
For simplicity, we consider the case when $V_{\beta}$ has nonzero elements only for a single orbital $\beta = \gamma$, and set $V_{\beta}^{\mathrm{eh}} = V_{\beta}^{\mathrm{he}}$ to be real. Under these assumptions, we set the notation.
\begin{equation}
    \cos \theta = \frac{\frac{1}{2}\left( V^{\mathrm{ee}}_{\gamma}\!-\!V^{\mathrm{hh}}_{\gamma} \right) }{ \sqrt{ \big[ \frac{1}{2}\left( V^{\mathrm{ee}}_{\gamma}\!-\!V^{\mathrm{hh}}_{\gamma} \right) \big]^2 \!+\! \left| V^{\mathrm{eh}}_{\gamma} \right|^2 } }, \quad 
    \sin \theta = \frac{ V^{\mathrm{eh}}_{\gamma} }{ \sqrt{ \big[ \frac{1}{2}\left( V^{\mathrm{ee}}_{\gamma}\!-\!V^{\mathrm{hh}}_{\gamma} \right) \big]^2 \!+\! \left| V^{\mathrm{eh}}_{\gamma} \right|^2 } }.
    \label{eq: pre notation}
\end{equation}
The problem of solving Eq.~\eqref{eq: SE1} is related to the eigenstates and eigenvalues of the operator $V_{\gamma}$. Thus for a given operator $V_\gamma$, there will be two solutions, one is electron-like and the other is hole-like.
    \begin{align}
        \text{Eigenstates:}&  \ \psi_{\alpha}^{\mathrm{(e/h)}}(x) = \frac{1}{\sqrt{\mathcal{N}}}\sum_{k} e^{ikx}u_{\alpha}(k) u^*_{\gamma}(k)
        \begin{pmatrix}
            \sqrt{\frac{1}{2}\left( 1 \!\pm\! \cos \theta \right)} e^{i\varphi/2} \\
            \pm \sqrt{\frac{1}{2}\left( 1 \!\mp\! \cos \theta \right)}\\
        \end{pmatrix},\\
        \text{Eigenvalues:}& \ E^{\mathrm{(e/h)}} = \frac{1}{N} \sum_{k}|u_{\gamma}(k)|^2\left[ \frac{1}{2}\left( V^{\mathrm{ee}}_{\gamma}\!+ \!V^{\mathrm{hh}}_{\gamma} \right) \!\pm \!\sqrt{ \big[ \frac{1}{2}\left( V^{\mathrm{ee}}_{\gamma}\!-\!V^{\mathrm{hh}}_{\gamma} \right) \big]^2 \!+\! \left| V^{\mathrm{eh}}_{\gamma} \right|^2 }\right],
        \label{eq: SE2}
    \end{align}
with $\mathcal{N}$ some normalization constant. At low temperatures, only the lower energy (hole-like) eigenstate contributes a large current. For conciseness, we only focus on the state $\psi_{\alpha}^{(\mathrm{h})}(x)$ and suppress all the labels $\mathrm{(e/h)}$ in Eq.~\eqref{eq: SE2}. We apply the notation $\psi_{\alpha}(x)$ to denote $\psi_{\alpha}^{\mathrm{(h)}}(x)$. In the following discussion, $\gamma= \mathrm{A} $ is used as an example to show the mechanism that the supercurrent is carried by the  interface states.

For a flat band Josephson junction, $V_{\mathrm{L}}$ and $V_{\mathrm{R}}$ would generate two interface states localized at the left and right interfaces, respectively. According to previous discussions, $V_{{\mathrm{L/R}}}$ would give rise to $4$ interface states, with left/right interfaces partners and electron-like/ hole-like partners. We use the notation $\ket{\psi_{\mathrm{L/R}}}$ and $E_{\mathrm{L/R}}^0$ to indicate the hole-like boundary states and their energies arising from $V_{\mathrm{L/R}}$. The interplay of these two states encourages us to set up a phenomenological model to describe how these bound states mediate supercurrent in a flat band Josephson junction. 

We can project the Hamiltonian in Eq.~\eqref{eq: JJ Ham} on the manifold spanned by $\ket{\psi_{\mathrm{L}}}$ and $\ket{\psi_{\mathrm{R}}}$. This is legitimate since the other states in the flat band are compact localized states with strictly zero amplitude on the boundary site. Thus, their energies would be insensitive to phase change. An effective $2\times2$ Hamiltonian can be obtained after the projection
\begin{equation}
    H = 
    \begin{pmatrix}
        E^0_{\mathrm{L}} + \bra{\psi_{\mathrm{L}}} V_{\mathrm{R}}\ket{\psi_{\mathrm{L}}} & \bra{\psi_{\mathrm{L}}}V_{\mathrm{L}} + V_{\mathrm{R}}\ket{\psi_{\mathrm{R}}}\\
        \bra{\psi_{\mathrm{R}}}V_{\mathrm{L}} + V_{\mathrm{R}}\ket{\psi_{\mathrm{L}}} & E^0_{\mathrm{R}} + \bra{\psi_{\mathrm{R}}} V_{\mathrm{L}}\ket{\psi_{\mathrm{R}}}\\
    \end{pmatrix}.
    \label{eq: effective H}
\end{equation}
The eigenenergies are
\begin{equation}
\begin{aligned}
    E_{\pm} = & \frac{E^0_{\mathrm{L}}+E^0_{\mathrm{R}}}{2} + \frac{\bra{\psi_{\mathrm{L}}} V_{\mathrm{R}}\ket{\psi_{\mathrm{L}}} + \bra{\psi_{\mathrm{R}}} V_{\mathrm{L}}\ket{\psi_{\mathrm{R}}}}{2} \\
    &\pm\sqrt{\Bigg( \frac{E^0_{\mathrm{L}}-E^0_{\mathrm{R}}}{2} + \frac{\bra{\psi_{\mathrm{L}}} V_{\mathrm{R}}\ket{\psi_{\mathrm{L}}} - \bra{\psi_{\mathrm{R}}} V_{\mathrm{L}}\ket{\psi_{\mathrm{R}}}}{2} \Bigg)^2 + |\bra{\psi_{\mathrm{L}}}V_{\mathrm{L}} + V_{\mathrm{R}}\ket{\psi_{\mathrm{R}}}|^2}.
\end{aligned}\label{eq: eigenenergy}
\end{equation}
These energies are sufficient to obtain the energy phase relation of the Josephson junction
\begin{equation}
    F(\varphi) = -\frac{1}{\beta}[\ln(1+e^{-\beta E_+}) + \ln(1+e^{-\beta E_-}) ]
\end{equation}
In the Lieb lattice case, the inversion symmetry is broken and a large splitting between $E^0_{\mathrm{L}}$ and $E^0_{\mathrm{R}}$ shows up. While the other terms in Hamiltonian Eq.~\eqref{eq: effective H} all fall exponentially as a function of the junction length $L$. Thus the relation $|E^0_{\mathrm{L}}-E^0_{\mathrm{R}}| \gg $ (other terms in $H$) holds. The square root term in Eq.~\eqref{eq: eigenenergy} can be expanded preserving first order term:
\begin{equation}
    E_{\pm} \approx  
    \begin{cases}
    E^0_{\mathrm{L}} + \bra{\psi_{\mathrm{L}}} V_{\mathrm{R}}\ket{\psi_{\mathrm{L}}} + \frac{2}{E^0_{\mathrm{L}}-E^0_{\mathrm{R}}}|\bra{\psi_{\mathrm{L}}}V_{\mathrm{L}} + V_{\mathrm{R}}\ket{\psi_{\mathrm{R}}}|^2 \\
    E^0_{\mathrm{R}} + \bra{\psi_{\mathrm{R}}} V_{\mathrm{L}}\ket{\psi_{\mathrm{R}}} - \frac{2}{E^0_{\mathrm{L}}-E^0_{\mathrm{R}}}|\bra{\psi_{\mathrm{L}}}V_{\mathrm{L}} + V_{\mathrm{R}}\ket{\psi_{\mathrm{R}}}|^2
    \end{cases}
\label{eq: eigenenergy approx}
\end{equation}
Substituting $\ket{\psi_{\mathrm{L/R}}}$ by full expressions in Eq.~\eqref{eq: SE2}, we find the last two terms share the form $A_0 + A_1 \cos \varphi$. As an example, this can be seen if the self-energy correction $V_{\mathrm{L}}$ and $V_{\mathrm{R}}$ have nonzero components only on the A site. We can apply the notation of $\psi_{\mathrm{A}}(x)$ in Eqs.~\eqref{psiA}-~\ref{psiC}:
\begin{align}
\label{eq: correction}
\bra{\psi_{\mathrm{L}}} V_{\mathrm{R}}\ket{\psi_{\mathrm{L}}} = &\ \frac{1}{\mathcal{N}_{\mathrm{L}}} |\psi_{\mathrm{A}}(L)|^2  \left[ (\varphi \text{ independent terms}) - |\sin \theta|V^{\mathrm{eh}} \cos\varphi \right]\notag\\
\bra{\psi_{\mathrm{R}}} V_{\mathrm{L}}\ket{\psi_{\mathrm{R}}} = &\ \frac{1}{\mathcal{N}_{\mathrm{R}}} |\psi_{\mathrm{A}}(-L)|^2  \left[(\varphi \text{ independent terms}) - |\sin \theta|V^{\mathrm{eh}} \cos\varphi \right]\\
|\bra{\psi_{\mathrm{L}}}V_{\mathrm{L}} + V_{\mathrm{R}}\ket{\psi_{\mathrm{R}}}|^2 = &\  \frac{1}{\mathcal{N}_{\mathrm{L}} \mathcal{N}_{\mathrm{R}}} \left|\psi_{\mathrm{A}}^*(L) \psi_{\mathrm{A}}(0) + \psi_{\mathrm{A}}^*(0) \psi_{\mathrm{A}}(-L) \right|^2 \left[ U + \frac{W}{2}\cos\varphi \right]\notag
\end{align}
with 
\begin{equation}
\begin{aligned}
\quad U =&\ \frac{1}{4}\left\{ [(1-\cos \theta)V^{\mathrm{ee}}]^2 + [(1+\cos \theta)V^{\mathrm{hh}}]^2 + 2|\sin \theta| V^{\mathrm{eh}}[(1-\cos \theta)V^{\mathrm{ee}} + (1+\cos \theta)V^{\mathrm{hh}}]\right\},\\
W =&\ \sin^2\theta (V^{\mathrm{ee}} V^{\mathrm{hh}} + (V^{\mathrm{eh}})^2) + |\sin \theta| V^{\mathrm{eh}}\left[ (1-\cos \theta) V^{\mathrm{ee}} + (1+\cos \theta) V^{\mathrm{hh}} \right].
\end{aligned}
\end{equation}
Here $\mathcal{N}_{\mathrm{L}}$ and $\mathcal{N}_{\mathrm{R}}$ are normalization constants of the left and right wavefunction, respectively, and the Josephson current can be obtained by $I =\frac{2e}{\hbar} \frac{\partial F(\varphi)}{\partial \varphi}$. If the $\varphi$-dependent part of the third term has a larger amplitude than the second term, one will observe that the eigenstate with lower energy will contribute a current proportional to $\sin \varphi$ (left interface state), while the one with higher energy will contribute a current with the phase shift $\pi$ (right interface state). Combining with Eq.~(\ref{eq: correction}), we define the amplitude of the $\varphi$-dependent part of the energy correction in Eq.~(\ref{eq: eigenenergy approx}) as: 
\begin{equation}
        E^1_\mathrm{L/R}=\left[\frac{-|\sin \theta|V^{\mathrm{eh}}}{\mathcal{N}_\mathrm{L/R}} 
        \pm\frac{4W}{\mathcal{N}_{\mathrm{L}} \mathcal{N}_{\mathrm{R}}(E^0_\mathrm{L}-E^0_\mathrm{R})}\right]e^{-4\sqrt{2}\delta L}.
\end{equation}


\begin{figure}[t]
    \centering
    \includegraphics[scale=1]{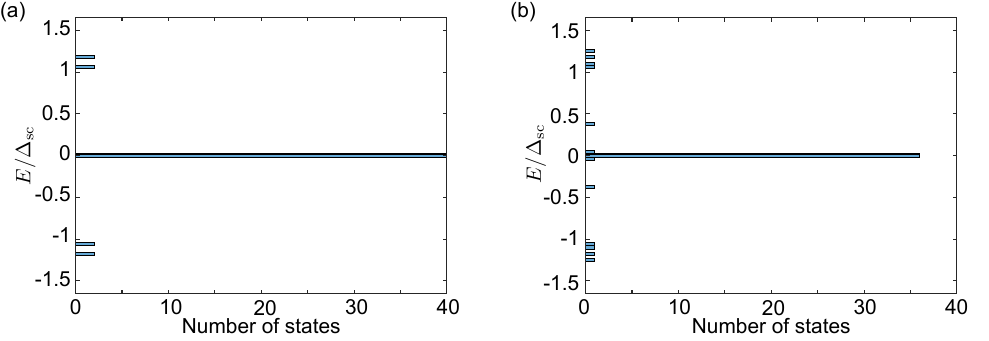}
    \caption{ Counting of numbers of states within a narrow energy window around the flat band energy $E=0$. Different coupling between weak link and superconducting lead: (a) $T_{\mathrm{l/rA}}=0$, and (b) $T_{\mathrm{l/rA}}=100 \Delta_{\mathrm{sc}}$. The $T_{\mathrm{l/rA}}=0$ case indicates an isolated weak link with no coupling with the superconducting lead, while for finite coupling $T_{\mathrm{l/rA}}=100\Delta_{\mathrm{sc}}$, four bound states show up in the spectrum. Here the parameters are $\Delta_{\mathrm{sc}}=1$, $J = 10^4\Delta_{\mathrm{sc}}$ and $t = 100\Delta_{\mathrm{sc}}$. The junction length is $L=20a$.}
    \label{FigS4}
\end{figure}


And the energies $E_{\pm}$ can be expressed as
\begin{equation}\label{approx E}
    E_{\pm} \approx E^0_\mathrm{L/R} + (\varphi\text{ independent part})e^{-4\sqrt{2}\delta L} + E^1_\mathrm{L/R}\cos \varphi + \mathcal{O}(e^{-8\sqrt{2}\delta L}) .
\end{equation}
In the Lieb lattice, we can simplify $f(E_{+/-})\approx f(E^0_\mathrm{L/R})$ since the last three terms in Eq.~(\ref{approx E}), suppressed by the exponential factor, are much smaller than $|E^0_\mathrm{L/R}|$. Here $f(E)$ is the Fermi-Dirac distribution. Besides, the flat band bound state on A site has the form given by Eq.~(\ref{psiA}). Thus the current phase relation is
\begin{equation}
\begin{aligned}\label{approx I}
    I(\varphi)&=\big(I_+f(E^0_\mathrm{L})-I_-f(E^0_\mathrm{R})\big)\sin\varphi, \\
              &\propto\exp{(-L/4\xi_{\mathrm{QM}})}\sin{\varphi},
\end{aligned}
\end{equation}
where $I_{+/-}=\frac{2e}{\hbar}|E^1_\mathrm{L/R}|$, and $\xi_{\mathrm{QM}}$ is the quantum metric length scale defined in Eq.~(\ref{QM_length}).
The supercurrent is mediated by these interface states, which decay exponentially into the weak link. In a flat-band case with an extremely large band gap, the multi-Andreev-reflection processes are totally suppressed, as such the current phase relation persists to be proportional to $\sin \varphi$ at low temperature.
In the Lieb lattice, because of the condition $E^0_\mathrm{L}<E^0_\mathrm{R}<0$ when the chemical potential is set at zero energy, both of the two states are occupied at zero temperature. With an increasing temperature, $f(E^0_\mathrm{R})$ decreases faster than $f(E^0_\mathrm{L})$, giving rise to a non-monotonic temperature dependence of the critical current at the low temperature regime, as shown in the main text.

We also calculate the density of states in the flat band Josephson junction [see Fig.~\ref{FigS4}]. The coupling between the weak link and the superconductor lead $T_{\mathrm{l/rA}}$ is set to $0$ for Fig.~\ref{FigS4}(a), and $T_{\mathrm{l/rA}}=100 \Delta_{\mathrm{sc}}$ for Fig.~\ref{FigS4}(b). When $T_{\mathrm{l/rA}}$ is turned on from 0, the energy of 4 states will be raised from the flat band energy, and give rise to the four bound states depicted in Fig.~\ref{FigS4}(b). We also estimate the bandwidth of the lowing flat-band states, which turns out to be significantly smaller than the order of critical current. This shows that the supercurrent in flat band Josephson junction is originated from the bound states.

\section{Current-Phase Relation from the Free Energy via Path Integral Method}

In this section, based on the lattice model $H_{\mathrm{JJ}}$ given in the main text, we apply the path integral method to derive an analytical formula for the current-phase relation. Spin degrees of freedom are always degenerate, thus the spin indices are not written explicitly in the following discussion. 
The Hamiltonian $H_{\mathrm{JJ}}$ can be reformulated in $k-$space as
\begin{align}
    \label{H_JJ_k}
    H_{\mathrm{JJ}}&=\sum_{k_\mathrm{l}}\epsilon_{k_\mathrm{l}}c^\dagger_{k_\mathrm{l}}c_{k_\mathrm{l}}+(\Delta e^{i\varphi_\mathrm{l}}c^\dagger_{k_\mathrm{l}}c^\dagger_{-k_\mathrm{l}}+\mathrm{h.c.})+\sum_{k_\mathrm{r}}\epsilon_{k_\mathrm{r}}c^\dagger_{k_\mathrm{r}}c_{k_\mathrm{r}}+(\Delta e^{i\varphi_\mathrm{r}}c^\dagger_{k_\mathrm{r}}c^\dagger_{-k_\mathrm{r}}+\mathrm{h.c.})\notag\\
    &+\frac{1}{\sqrt{N_{\mathrm{l}}N}}\big(\sum_{k_\mathrm{l}k}\sum_{\alpha}T_{\mathrm{l}\alpha}c^\dagger_{k_\mathrm{l}}a_{\alpha k}+T^*_{\mathrm{l}\alpha}a^\dagger_{\alpha k}c_{k_\mathrm{l}}\big)+\frac{1}{\sqrt{N_{\mathrm{r}}N}}\sum_{k_\mathrm{r}k}\sum_{\alpha}\big(T_{\mathrm{r}\alpha}c^\dagger_{k_\mathrm{r}}a_{\alpha k}e^{ikL}+T^*_{\mathrm{r}\alpha}a^\dagger_{\alpha k}c_{k_\mathrm{r}}e^{-ikL}\big)\notag \\
    &+\sum_{k}\sum_{\alpha\beta}h_{\alpha\beta}(k)\delta_{\alpha\beta}a^\dagger_{\alpha k}a_{\beta k},
\end{align}
where $c_{k_\mathrm{{l/r}}}$ ($c^\dagger_{k_\mathrm{l/r}}$) are the annihilation (creation) operators of the electrons with momentum $k_{\mathrm{l/r}}$ in the left/right SC lead with pairing potential $\Delta e^{i\varphi_{\mathrm{l/r}}}$,  $a_k$ ($a^\dagger_k$) are the annihilation (creation) operators of the electrons with momentum $k$ in Lieb lattice, the matrix $h(k)$ has the form given by Eq.~(\ref{Lieb_k}), and $T_{\mathrm{l/r}\alpha}$ are the hopping amplitudes from the atomic orbital $\alpha$ at the leftmost/rightmost site of Lieb lattice to the left/right SC lead. Besides, $N_{\mathrm{l/r}}$ are the number of site of the left/right lead respectively, and $N$ is the number of site of the flat band weak link.  We will derive the current-phase relation without loss of generality by setting the chemical potential at the Lieb lattice as $\mu_{\mathrm{N}}=0$, which coincides with the flat-band energy. 

Due to the isolatedness of the flat band, we can project Lieb lattice on the subspace spanned by the Bloch state of the flat band $\{u_{\bar n k}\}$,
\begin{equation}
a_{\alpha k}\rightarrow u^*_{\Bar{n}\alpha}(k)a_{\Bar{n}k}
\end{equation}
where $u_{\Bar{n}\alpha}(k)$ is the $\alpha$ component of $|u_0(k)\rangle$ defined in Eq.~(\ref{Lieb_k}). After projection, the contact Hamiltonian is
\begin{align}  H_\mathrm{c}&=\frac{1}{\sqrt{N_{\mathrm{l}}N}}\sum_{k_\mathrm{l}k}\sum_{\alpha}T_{\mathrm{l}\alpha}c^\dagger_{k_\mathrm{l}}u^*_{\Bar{n}\alpha}(k)a_{\Bar{n}k}+\frac{1}{\sqrt{N_{\mathrm{r}}N}}\sum_{k_\mathrm{r}k}\sum_{\alpha}T_{\mathrm{r}\alpha}c^\dagger_{k_\mathrm{r}}u^*_{\Bar{n}\alpha}(k)a_{\Bar{n}k}e^{ikL}+\mathrm{h.c.} \notag \\
&=\sum_{k_\mathrm{l}k}T^{k_\mathrm{l}k}_{\mathrm{l}\Bar{n}}c^\dagger_{k_\mathrm{l}}a_{\Bar{n}k}+\sum_{k_\mathrm{r}k}T^{k_\mathrm{r}k}_{\mathrm{r}\Bar{n}}c^\dagger_{k_\mathrm{r}}a_{\Bar{n}k}e^{ikL}+\mathrm{h.c.} ,
\end{align}
where $T^{k_{\mathrm{l/r}}k}_{\mathrm{l/r}\Bar{n}}=\frac{1}{\sqrt{N_{\mathrm{l/r}}N}}\sum_{\alpha}T_{\mathrm{l/r}\alpha}u^*_{\Bar{n}\alpha}(k)$.
With the corresponding Grassmann variables, $c_{k_\mathrm{l/r}}\rightarrow\psi_{k_\mathrm{l/r}}$, $c^\dagger_{k_\mathrm{l/r}}\rightarrow\Bar{\psi}_{k_\mathrm{l/r}}$, $a_{\Bar{n} k}\rightarrow \psi_{\Bar{n}k}$ and $a^\dagger_{\Bar{n} k}\rightarrow \Bar{\psi}_{\Bar{n}k}$,
we have the partition function $Z(\varphi)= \mathrm{Tr} e^{-\beta H_{\mathrm{JJ}}}$ with $\varphi=\varphi_\mathrm{l}-\varphi_\mathrm{r}$  and in details,
\begin{align}
    Z(\varphi)&=\int D[\psi_{k_\mathrm{l}},\Bar{\psi}_{k_\mathrm{l}},\psi_k,\Bar{\psi}_k,\psi_{k_\mathrm{r}},\Bar{\psi}_{k_\mathrm{r}}]\exp{(-S)}\\
    S&=\int_0^\beta d\tau\sum_{k_\mathrm{l}}\Bar{\psi}_{k_\mathrm{l}}(\partial_\tau+\epsilon_{k_\mathrm{l}})\psi_{k_\mathrm{l}}+(\Delta e^{i\varphi_\mathrm{l}} \Bar{\psi}_{k_\mathrm{l}}\Bar{\psi}_{-k_\mathrm{l}}+h.c.)+\sum_{k_\mathrm{r}}\Bar{\psi}_{k_\mathrm{r}}(\partial_\tau+\epsilon_{k_\mathrm{r}})\psi_{k_\mathrm{r}}+(\Delta e^{i\varphi_\mathrm{r}}\Bar{\psi}_{k_\mathrm{r}}\Bar{\psi}_{-k_\mathrm{r}}+\mathrm{h.c.}) \notag \\
    &+\Bar{\psi}_{\Bar{n}k}\partial_\tau\psi_{\Bar{n}k}+\sum_{k_\mathrm{l}k}T^{k_\mathrm{l}k}_{\mathrm{l}\Bar{n}}\Bar{\psi}_{k_\mathrm{l}}\psi_{\Bar{n}k}+\sum_{k_\mathrm{r}k}T^{k_\mathrm{r}k}_{\mathrm{r}\Bar{n}}\Bar{\psi}_{k_\mathrm{r}}\psi_{\Bar{n}k}e^{ikL}+\mathrm{h.c.},
\end{align}
where $\beta=1/k_{\mathrm{B}}T$. By integrating out $\psi_{\bar n k }$ in Lieb lattice, the action can be reduced to an effective form of the well-known tunneling junction, and by using the Fourier transform $\psi_{k_\mathrm{l/r}}(\tau)=\frac{1}{\sqrt{\beta}}\sum_{\omega_m}\psi_{k_\mathrm{l/r}m}e^{-i\omega_m\tau}$, where $\omega_m=(2m+1)\pi/\beta$ is Matsubara frequency, the effective action in frequency basis can be written as
\begin{align}
    S_{\mathrm{eff}}&= \sum_{\omega_m}\sum_{k_\mathrm{l}}\Bar{\psi}_{k_\mathrm{l}m}(-i\omega_m+\epsilon_{k_\mathrm{l}}+\Sigma_\mathrm{l}(\omega_m))\psi_{k_\mathrm{l}m}+(\Delta e^{i\varphi_l}\Bar{\psi}_{k_\mathrm{l}m}\Bar{\psi}_{-k_\mathrm{l}-m}+\mathrm{h.c.}) \notag\\
    & +  \sum_{\omega_m}\sum_{k_\mathrm{r}}\Bar{\psi}_{k_\mathrm{r}m}(-i\omega_m+\epsilon_{k_\mathrm{r}}+\Sigma_\mathrm{r}(\omega_m))\psi_{k_\mathrm{r}m}+(\Delta e^{i\varphi_\mathrm{r}}\Bar{\psi}_{k_\mathrm{r}m}\Bar{\psi}_{-k_\mathrm{r}-m}+\mathrm{h.c.}) \notag \\
&+\sum_{\omega_m}\sum_{k_\mathrm{l}k_\mathrm{r}}V^{\mathrm{ee}}_{\mathrm{lr}}(\omega_m)\Bar{\psi}_{k_\mathrm{l}m}\psi_{k_\mathrm{r}m}+V^{\mathrm{ee}}_{\mathrm{rl}}(\omega_m)\Bar{\psi}_{k_\mathrm{r}m}\psi_{k_\mathrm{l}m}
\end{align}
The real-space Green's function of the flat band system is defined as $G^{\mathrm{FB}}_{\alpha\beta}(\omega_m,x)=\frac{1}{N}\sum_{k}u_{\alpha}(k)u^*_{\beta}(k)e^{ikx}\frac{1}{-i\omega_m}$, by which the electron part of the effective coupling strengths between SC leads can be expressed as $V^{\mathrm{ee}}_{\mathrm{lr}}(\omega_m)=\frac{1}{\sqrt{N_{\mathrm{l}}N_{\mathrm{r}}}}\sum_{\alpha\beta}T^*_{\mathrm{r}\alpha}T_{\mathrm{l}\beta}G^{\mathrm{FB}}_{\alpha\beta}(\omega_m, -L)$, where $L$ is the length of the junction, $V^{\mathrm{ee}}_{\mathrm{rl}}(\omega_m)=\frac{1}{\sqrt{N_{\mathrm{l}}N_{\mathrm{r}}}}\sum_{\alpha\beta}T_{\mathrm{r}\alpha}T^*_{\mathrm{l}\beta}G^{\mathrm{FB}}_{\alpha\beta}(\omega_m, L)$, and $\Sigma_{\mathrm{l/r}}(\omega_m)=\frac{1}{N_{\mathrm{l/r}}}\sum_{\alpha\beta}T_{\mathrm{l/r}\alpha}T^*_{\mathrm{l/r}\beta}G^{\mathrm{FB}}_{\alpha\beta}(\omega_m, 0)$ is the self-energy correction of the left/right SC lead induced by the coupling with Lieb lattice. As an approximation, only the diagonal elements of the self-energy correction are taken into account. The length-dependent part of the effective coupling strength is the same as the bound state wavefunction derived in section~\ref{Analytical Model}. The Gor'kov Green's function of the left/right SC lead is 
\begin{equation}
    G_{\mathrm{l/r}}(\omega_m,k_\mathrm{{l/r}})=\begin{pmatrix}
        G^\mathrm{ee}_{\mathrm{l/r}} & G^\mathrm{eh}_{\mathrm{l/r}}\\
        G^\mathrm{he}_{\mathrm{l/r}} & G^\mathrm{hh}_{\mathrm{l/r}}
    \end{pmatrix}(\omega_m,k_\mathrm{{l/r}})=\begin{pmatrix}
        -i\omega_m+\epsilon_{k_\mathrm{{l/r}}}+\Sigma_{\mathrm{l/r}}(\omega_m) & \Delta e^{i\varphi_{\mathrm{l/r}}}\\
        \Delta e^{-i\varphi_{\mathrm{l/r}}} & -i\omega_m-\epsilon_{k_\mathrm{l/r}}-\Sigma_{\mathrm{l/r}}(\omega_m)^*
    \end{pmatrix}^{-1}.
\end{equation}
After integrating out the degrees of freedom in SC leads, the partition function $Z(\varphi)$ of the flat band Josephson junction and thus the free energy $F(\varphi)$ can be obtained. It is the $\varphi$-dependent part that is of significance for Josephson effect. The Green's function of the two uncoupled SC leads is defined as $\hat{G}=\begin{pmatrix}
    G_\mathrm{l} & 0\\ 0 & G_\mathrm{r}
\end{pmatrix}$, and the matrix of coupling strength is $\hat{V}=\begin{pmatrix}
    0 & V_{\mathrm{lr}}\\V_{\mathrm{rl}} & 0
\end{pmatrix}$, where $V_{\mathrm{lr}/\mathrm{rl}}=\begin{pmatrix}
    V^\mathrm{ee}_{\mathrm{lr}/\mathrm{rl}} & 0\\ 0 & V^\mathrm{hh}_{\mathrm{lr}/\mathrm{rl}}
\end{pmatrix}$ and $V^\mathrm{hh}_{\mathrm{lr/\mathrm{rl}}}=-(V^\mathrm{ee}_{\mathrm{lr}/\mathrm{rl}})^*$. The free energy can be expressed in terms of $\hat{G}$ and $\hat{V}$ as
\begin{align}
    F(\varphi)&=-\frac{1}{\beta}\ln(Z(\varphi)) \notag\\
    &=-\frac{1}{\beta}\mathrm{tr}\ln\beta\begin{pmatrix}
        G^{-1}_\mathrm{l} & V\\
        V^\dagger & G^{-1}_\mathrm{r}
    \end{pmatrix}\notag\\
    &=-\frac{1}{\beta}\mathrm{tr}\ln(\beta\hat{G}^{-1}(1+\hat{G}\hat{V}))
\end{align}
We further assume small coupling condition that $|\sum_{\alpha\beta}T_{\mathrm{r}\alpha}T^*_{\mathrm{l}\beta}\sum_{\mathrm{k}}u_{\alpha}(k)u^*_{\beta}(k)e^{ikL}|\ll|t|^2$, where $t$ is the hopping amplitude in SC lead. Then, by using the relation $I(\varphi)=\frac{2e}{\hbar}\frac{\partial F}{\partial\varphi}$, up to the lowest order, the current-phase relation can be calculated by
\begin{align}
\label{CPR}
    I(\varphi)&=\frac{2e}{\hbar\beta}\mathrm{tr}\frac{\partial}{\partial\varphi}\big(\frac{1}{2}\hat{G}\hat{V}\hat{G}\hat{V}\big) \notag\\
    &=\frac{2e}{\hbar\beta}\mathrm{tr}\frac{\partial}{\partial\varphi}
    \frac{1}{2}\begin{pmatrix}
        G_\mathrm{l}V_{\mathrm{lr}}G_\mathrm{r}V_{\mathrm{rl}} & 0\\
        0 & G_\mathrm{r}V_{\mathrm{rl}} G_\mathrm{l}V_{\mathrm{lr}}
    \end{pmatrix}. 
\end{align}

\begin{figure}[t]
    \centering
    \includegraphics[scale=0.85]{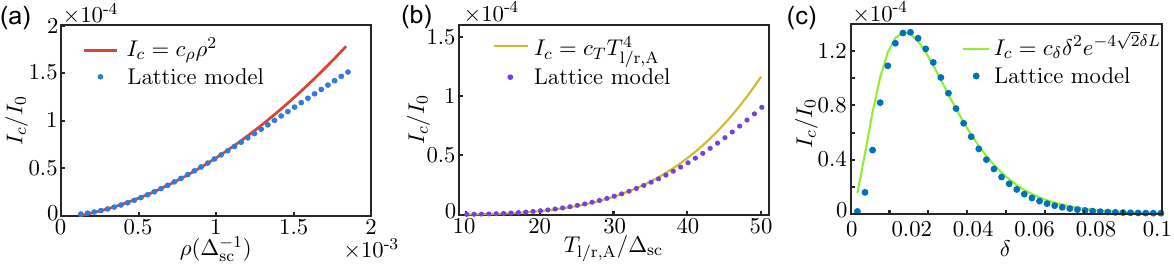}
    \caption{ Critical current as a function of (a) the electronic density of state (DoS) $\rho$ at the Fermi surface of the leads; (b) coupling strength $T_{\mathrm{l/r,A}}$; (c) parameter $\delta$. (a) For a small density of states, the critical current scales as $I_c\propto \rho^2$. (b) For a small coupling strength, critical current scales as $I_c\propto|T_{\mathrm{A}}|^4$. (c) The critical current behaves nonmonotically when $\delta$ is changed. And the maximum value shows up at $\delta=1/(2\sqrt{2}L)$. These figures are calculated for $k_{\mathrm{B}}T = 0.1\Delta_{\mathrm{sc}}$. Especially, $\delta=0.03$, $T_{\mathrm{l/r,A}}=30 \Delta_{\mathrm{sc}}$, $L=20 a$ for (a) , $\delta=0.03$, $t=100 \Delta_{\mathrm{sc}}$, $L=20 a$ for (b) and $T_{\mathrm{l/r,A}}=30 \Delta_{\mathrm{sc}}$, $t=70 \Delta_{\mathrm{sc}}$, $L=50 a$ for (c).}
    \label{Path_int}
\end{figure}

Now each part in Eq.~(\ref{CPR}) can be calculated separately to give the total current phase relation. To exemplify the method, we assume that $T_{\alpha}=T_\mathrm{A}\delta_{\mathrm{A}\alpha}$, and the two SC leads are the same except for the phase difference. 
\begin{align}
    &\frac{1}{\beta}\mathrm{tr}G_\mathrm{l}V_{\mathrm{lr}}G_\mathrm{r}V_{\mathrm{rl}} \notag \\
    =&\frac{1}{\beta}\mathrm{tr}\begin{pmatrix}
        G^\mathrm{eh}_\mathrm{l}V_{\mathrm{lr}}^\mathrm{hh}G^\mathrm{he}_\mathrm{r}V_{\mathrm{rl}}^\mathrm{ee} & 0\\
        0 & G^\mathrm{he}_\mathrm{l}V_{\mathrm{lr}}^\mathrm{ee}G^\mathrm{eh}_\mathrm{r}V_{\mathrm{rl}}^\mathrm{hh}
    \end{pmatrix} \notag\\
    =&\frac{1}{\beta N_{\mathrm{l}}N_{\mathrm{r}}}\sum_{\omega_m}\sum_{k_\mathrm{l}k_\mathrm{r}}\frac{1}{2}\big(\frac{\Delta e^{i\varphi_\mathrm{l}}}{(-i\omega_m+\epsilon_{k_\mathrm{l}}+\Sigma_l)(-i\omega_m-\epsilon_{k_\mathrm{l}}-\Sigma_\mathrm{l}^*)-\Delta^2}\frac{\Delta e^{-i\varphi_\mathrm{r}}}{(-i\omega_m+\epsilon_{k_\mathrm{r}}+\Sigma_\mathrm{r})(-i\omega_m-\epsilon_{k_\mathrm{r}}-\Sigma_\mathrm{r}^*)-\Delta^2}\frac{|T_A|^4\delta^2 e^{-4\sqrt{2}\delta L}}{(-i\omega_m)^2}\notag\\
    +&\frac{\Delta e^{-i\varphi_\mathrm{l}}}{(-i\omega_m+\epsilon_{k_\mathrm{l}}+\Sigma_\mathrm{l})(-i\omega_m-\epsilon_{k_\mathrm{l}}-\Sigma_\mathrm{l}^*)-\Delta^2}\frac{\Delta e^{i\varphi_\mathrm{r}}}{(-i\omega_m+\epsilon_{k_\mathrm{r}}+\Sigma_\mathrm{r})(-i\omega_m-\epsilon_{k_\mathrm{r}}-\Sigma_\mathrm{r}^*)-\Delta^2}\frac{|T_\mathrm{A}|^4\delta^2 e^{-4\sqrt{2}\delta L}}{(-i\omega_m)^2}\big)\notag\\
    =&-f(T_\mathrm{A},\delta,\beta,\Delta)\rho^2|T_\mathrm{A}|^4\delta^2 e^{-4\sqrt{2}\delta L}\cos(\varphi),
\end{align}
where $\rho$ is the electronic density of states of SC lead at Fermi energy, and the model-dependent coefficient $f(T_\mathrm{A},\delta,\beta,\Delta)$ can be expressed as 
\begin{equation}
    f(T_\mathrm{A},\delta,\beta,\Delta)=\frac{1}{\beta}\sum_{\omega_m}\int d\epsilon\int d\epsilon'\frac{\Delta}{(-i\omega_m\!+\!\epsilon\!+\!\Sigma_\mathrm{l})(-i\omega_m\!-\!\epsilon\!-\!\Sigma_\mathrm{l}^*)\!-\!\Delta^2}\frac{\Delta}{(-i\omega_m\!+\!\epsilon'\!+\!\Sigma_\mathrm{r})(-i\omega_m\!-\!\epsilon'\!-\!\Sigma_\mathrm{r}^*)\!-\!\Delta^2}\frac{1}{-(-i\omega_m)^2}.
\end{equation}
The assumed small coupling condition $|T_\mathrm{A}|^2\delta e^{-2\sqrt{2}\delta L}\ll |t|^2$ enables the approximation of the coefficient $f(T_\mathrm{A},\delta,T,\Delta)$ as independent of $\delta$ and $T_\mathrm{A}$. Thus the current-phase relation is
\begin{equation}\label{final}
    I(\varphi)=\frac{2e}{\hbar}f(\beta,\Delta)\rho^2|T_\mathrm{A}|^4\delta^2 e^{-4\sqrt{2}\delta L}\sin(\varphi).
\end{equation}
The critical current $I_c$ is predicted to satisfy: $I_c\propto \rho^2$, $I_c\propto|T_\mathrm{A}|^4$ and $I_c\propto \delta^2 e^{-4\sqrt{2}\delta L}$, which are consistent with the numerical results obtained from lattice model, as shown in Fig.~\ref{Path_int}. The length-dependence property of $I_c$ is determined by $e^{-4\sqrt{2}\delta L}$, which can be predicted by the localization length of the flat band bound state derived in section~\ref{Analytical Model}. Besides, since the A-component $u_{0\mathrm{A}}$ of flat band approaches zero as $\delta\rightarrow 0$, the $\delta$-dependence of $I_c$ is not monotonic, as shown in Fig.~\ref{Path_int}, which can be explained by the $\delta$-dependent part $\delta^2 e^{-4\sqrt{2}\delta L}$ in Eq.~\ref{final}. The maximum critical current is achieved at $\delta=1/(2\sqrt{2}L)$, where the localization length of the flat band bound state equals to the length of the junction. It further demonstrates the significance of the length scale defined by quantum metric in a flat band system.

\section{Josephson Current and the Lattice Green's Functions}
In this section, we explain how to use the lattice Green's function method to calculate the supercurrent transported through the flat-band Josephson junction depicted in the main text. The Hamiltonian of the Josephson junction can be divided into four parts as $H_{\mathrm{JJ}}=H_{\mathrm{R}}+H_{\mathrm{L}}+H_\mathrm{c}+H_{\mathrm{Lieb}}$, where $H_{\mathrm{R/L}}$ is the Hamiltonian of right/left superconducting lead, $H_\mathrm{Lieb}$ is the Hamiltonian of the central device (Lieb lattice), and $H_\mathrm{c}$ describes the coupling terms between the central device and the two sides superconducting leads. Given the Hamiltonian of the Josephson junction, many observables, especially such as the supercurrent can be expressed in terms of its Green's function.

The Green's function of the Josephson junction is defined as $G(i\omega_m)=(i\omega_m-H_{\mathrm{JJ}})^{-1}$, where $\omega_m=(2m+1)\pi k_\mathrm{B}T$ is the Matsubara frequency. However, it is not computationally efficient to calculate $G(i\omega_m)$ directly by its definition. Therefore, we choose the following approach:
\begin{itemize}
    \item  calculate the Green's function of each decoupled part of the Josephson junction separately;
    \item construct the Green's function of the entire coupled system in a recursive manner.
\end{itemize}
To facilitate the discussion, we define the Hamiltonian of an isolated unit cell of the central device:
\begin{equation}
H_{\mathrm{uc}}=\begin{pmatrix}
    H_{\mathrm{ee}} & 0 \\
    0 & H_{\mathrm{hh}}
\end{pmatrix},\ \text{with} \ H_{\mathrm{hh}}=-H^*_{\mathrm{ee}}.
\end{equation}
Here $H_{\mathrm{ee}}$ and $H_{\mathrm{hh}}$ are the electron and hole parts of the Hamiltonian, respectively. 
Similarly, the electron hopping amplitude from orbital $\beta$ at site $n+1$ to orbital $\alpha$ at site $n$ is denoted by $[W_\mathrm{ee}]_{\alpha\beta}$. Here we define 
\begin{equation}
W_\mathrm{L}=\begin{pmatrix}
    W_\mathrm{ee} & 0 \\
    0 & W_\mathrm{hh}
\end{pmatrix},\ \text{with} \ W_{\mathrm{hh}}=-W^*_\mathrm{ee}\  \text{and} \ W_\mathrm{R}=W^\dagger_\mathrm{L}.
\end{equation}
For Lieb lattice, in atomic orbital basis ($\alpha=\mathrm{A,B,C}$), the matrices can be written explicitly as 
\begin{align}
    H_{\mathrm{ee}}&=\begin{pmatrix}
        0 & J_+ & 0\\
        J_+ & 0 & J_0\\
        0 & J_0 & 0
    \end{pmatrix},\quad 
    W_{\mathrm{ee}}=\begin{pmatrix}
        0 & J_- & 0 \\
        0 & 0 & 0 \\
        0 & 0 & 0
    \end{pmatrix},
\end{align}
where $J_{\pm}=J(1\pm\delta)$ and $J_0=J\delta$. The Lieb lattice is assumed to be connected to the first site of each superconducting lead, so only the surface Green's function $g_\mathrm{L/R}$ of the left/right superconducting lead is needed. For a semi-infinite superconducting lead, the sites are labelled from $n_{\mathrm{sc,L/R}}=0$ to infinity, and thus the surface Green's function of the superconducting lead is defined as $g_{\mathrm{L/R}}=\langle n_{\mathrm{sc,L/R}}=0|G_{\mathrm{sc},\mathrm{L/R}}|n_{\mathrm{sc,L/R}}=0\rangle$, where $G_{\mathrm{sc},\mathrm{L/R}}=(i\omega-H_{\mathrm{L/R}})^{-1}$ is the Green's function of the whole left/right superconducting lead. Although the lead is semi-infinite, the surface Green's function can be calculated recursively without being bothered by the infinite dimensional Hamiltonian of the whole superconducting lead. Because the superconducting lead is chosen to be square lattice, the calculation of surface Green's function is standard, so we omit it at here. The hopping amplitude from the electron orbital $\alpha$ at the leftmost/rightmost site in Lieb lattice to the orbital at the first site in the left/right superconducting lead is $[T_{\mathrm{L/R},\mathrm{ee}}]_{\alpha}$, while the corresponding hole part is $[T_{\mathrm{L/R},\mathrm{hh}}]_{\alpha}=-[T_\mathrm{\mathrm{L/R},{ee}}]^*_{\alpha}$. The self-energy given by the left/right superconducting lead can be expressed as $\Sigma_\mathrm{L/R}=\begin{pmatrix}
    T^\dagger_{\mathrm{L/R},\mathrm{ee}} & 0\\
    0 & T^\dagger_{\mathrm{L/R},\mathrm{hh}}
\end{pmatrix}\begin{pmatrix}
    g_\mathrm{L/R,ee} & g_\mathrm{L/R,eh}\\
    g_\mathrm{L/R,he} & g_\mathrm{L/R, hh}
\end{pmatrix}\begin{pmatrix}
    T_{\mathrm{L/R},\mathrm{ee}} & 0 \\
    0 & T_{\mathrm{L/R},\mathrm{hh}}
\end{pmatrix}.$
In our calculation, we use $T_{\mathrm{L/R},\mathrm{ee}}=\begin{pmatrix}
    T_\mathrm{A} & 0 & 0
\end{pmatrix}$ as an example, where $T_\mathrm{A}$ depicts the coupling strength between Lieb lattice and superconducting lead through A site. Then Green's function $G(i\omega_m)$ of the whole Josephson junction can be constructed by the standard recursive method. Starting from $\tilde{G}_{n_\mathrm{l},n_\mathrm{l}}(i\omega_m)=(i\omega_m-H_{\mathrm{uc}}-\Sigma_\mathrm{L})^{-1}$ and $\tilde{G}_{n_\mathrm{r},n_\mathrm{r}}(i\omega_m)=(i\omega_m-H_{\mathrm{uc}}-\Sigma_\mathrm{R})^{-1}$, where $n_\mathrm{l}=1$, $n_\mathrm{r}=N$, and $N$ is the number of columns of the central device. The influence of the superconducting leads on the central device (Lieb lattice) can be included by recursion:  
    \begin{align}
        \Sigma_\mathrm{L/R}&=W_\mathrm{R/L}\tilde{G}_{n_\mathrm{l},n_\mathrm{l}/n_\mathrm{r},n_\mathrm{r}}W_\mathrm{L/R},\\
        \tilde{G}_{n_\mathrm{l}+1,n_\mathrm{l}+1}&=(i\omega_m-H_{\mathrm{uc}}-\Sigma_\mathrm{L})^{-1},\ \tilde{G}_{n_\mathrm{r}-1,n_\mathrm{r}-1}=(i\omega_m-H_{\mathrm{uc}}-\Sigma_\mathrm{R})^{-1}\\
        n_\mathrm{l}=n_\mathrm{l}+1&, \ n_\mathrm{r}=n_\mathrm{r}-1,
    \end{align}
until $n_\mathrm{l}=n_\mathrm{r}-1=n$, where $n$ corresponds to a site at the middle of the central device. Finally, the needed Green's function is $G_{n,n+1}$ ( $G_{n+1,n}$), where $[G_{n,n+1}]_{\alpha\beta}=\langle n\alpha|G(i\omega_n)|n+1,\beta\rangle$ is the matrix element of $G(i\omega_m)$ in the orbital basis $\{|n\alpha\rangle\}$ ($n$ labels site and $\alpha,\beta$ labels atomic orbital). It can be obtained from 
\begin{equation}
    \begin{pmatrix}
        G_{nn} & G_{n,n+1}\\
        G_{n+1,n} & G_{n+1,n+1}
    \end{pmatrix}=\begin{pmatrix}
        \tilde{G}^{-1}_{n_\mathrm{l}, n_\mathrm{l}} & W_\mathrm{L}\\
        W_\mathrm{R} & \tilde{G}^{-1}_{n_\mathrm{r}, n_\mathrm{r}}
    \end{pmatrix}^{-1}.
\end{equation}


\begin{figure}[ht]
    \centering
    \includegraphics[scale=1]{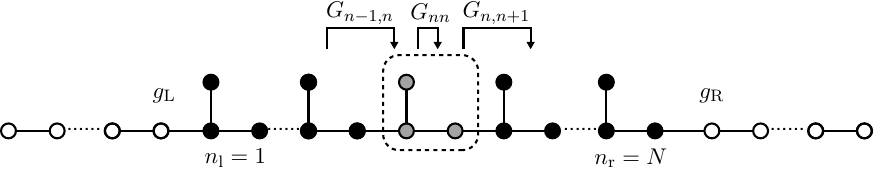}
    \caption{ Illustration for the algorithm of lattice Green's function method. We first obtained the surface Green's function $g_{L/R}$ from the superconducting leads. Subsequently, we employ an iterative method, commencing with the $n_{\mathrm{l}}=1$ and $n_{\mathrm{r}}=N$ sites.}
    \label{FigS5}
\end{figure}


Provided $G(i\omega_m)$, the supercurrent transported through the Josephson junction can be calculated by
\begin{equation}
I(\varphi) = \frac{2e}{\hbar}k_\mathrm{B}T\sum_{\omega_m}\mathrm{Im}\mathrm{Tr}(W_{n,n+1}G_{n+1,n}(i\omega_m)-W_{n+1,n}G_{n,n+1}(i\omega_m)),
\end{equation}
where $[W_{n,n+1}]_{\alpha\beta}$ includes the hopping amplitude between neighboring orbitals with matrix form
\begin{equation}
    W_{n,n+1}=\begin{pmatrix}
    W_\mathrm{ee} & 0\\
    0 & -W_\mathrm{hh}
\end{pmatrix},
\end{equation}
and $W_{n+1,n}=W^\dagger_{n,n+1}$.

Note that the Green's function $G_{n,n}$ at each site has 6 components, with orbital index $\alpha=\mathrm{A,B,C}$ and the particle-hole components. At zero temperature, the pairing correlation $|F(x)| = |\sum_{\alpha} \langle a_{x\alpha\uparrow}^{\dagger} a_{x\alpha\downarrow}^{\dagger} \rangle|$ can be obtained by taking the trace of anomalous Green's function 
\begin{equation}
    F(x)=\mathrm{Tr_o} \int \mathrm{d}E\ G_{x, x}^{(\mathrm{eh})}(E).
\end{equation}
Here $\mathrm{Tr_o}$ implies taking the trace over orbital indices, and $G_{x, x}^{(\mathrm{eh})}(E)$ means the anomalous Green's function at position $x$. For finite temperature case, the integral over $E$ is replaced by summing over the imaginary time Green's function $G_{x,x}^{(\mathrm{eh})}(i\omega_m)$ for Matsubara frequency $i\omega_m$.

\section{2D SC/FB/SC Josephson Junctions}
In this section, we present more details on the SC/flat band/SC (SC/FB/SC) Josephson junction and extend the discussions to two spatial dimensions. For the sake of convenience, we first present the model Hamiltonian. The total Hamiltonian describes the SC/FB/SC Josephson junction can be split into several parts 
\begin{align}
H_{\mathrm{JJ}}=H_{\mathrm{R}}+H_{\mathrm{L}}+H_{\mathrm{Lieb}}+H_{c}.
\end{align}
For two superconducting leads, we use $c_i$ and $c^\dagger_i$ to denote the creation and annihilation operators, respectively, at the site $i$, then $H_{\mathrm{sc}}$ can be written as
\begin{equation}\label{SC Ham}
    H_{\mathrm{sc}}=\sum_{\langle ij\rangle}(t-\mu_{\mathrm{s}}\delta_{ij})c^\dagger_ic_j+\sum_i(\Delta_{\mathrm{i}}c^\dagger_ic^{\dagger}_i+h.c.) 
\end{equation}
where $\langle ij\rangle$ means summing over nearest neighbor sites, $\Delta_{i}=\Delta_{\mathrm{sc}}e^{i\varphi_{\mathrm{l(r)}}}$ is the s-wave superconducting order parameter with phase $\varphi_\mathrm{l}$ and $\varphi_\mathrm{r}$ in left and right superconductors respectively, and $\mu_{\mathrm{s}}$ is the chemical potential of superconductor. Furthermore, the weak-link Hamiltonian can be expressed as
\begin{align}         
\label{Lieb tb Ham}
    H_{\mathrm{Lieb}}&=J(1+\delta)\sum_i(a^\dagger_{i\mathrm{A}}a^{}_{i\mathrm{B}}+a^\dagger_{i\mathrm{C}}a^{}_{i\mathrm{B}} + \mathrm{h.c.})\notag\\
    &+J(1-\delta)\sum_{i=(x,y)}(a^\dagger_{x-1,y,\mathrm{A}}a^{}_{x,y,\mathrm{B}} + a^\dagger_{x,y-1,\mathrm{C}}a^{}_{x,y,\mathrm{B}}
    + \mathrm{h.c.})\notag\\
    &-\mu_{\mathrm{N}}\sum_{i\alpha}a^\dagger_{i\alpha}a^{}_{i\alpha}.
\end{align}
while the coupling term is expressed as
\begin{equation}\label{coupling Ham}
    H_{\mathrm{c}}=\sum_{\alpha y}(T_{\mathrm{l}\alpha}c^\dagger_{0y} a_{1y\alpha}+T_{\mathrm{r}\alpha}c^\dagger_{N+1,y} a_{Ny\alpha} + \mathrm{h.c.}),
\end{equation}
where $a^{}_{i\alpha}$ ($a^\dagger_{i\alpha}$) is the creation (annihilation) operator of an electron on orbital $\alpha$ in a unit cell $i=(x,y)$ of Lieb lattice with $N$ sites along the direction of junction (x-axis), and $T_{\mathrm{l}\alpha}$ ($T_{\mathrm{r}\alpha}$) is the hopping amplitude from orbital $\alpha$ at the ends of the Lieb lattice to the first site of left (right) superconductor.

In two-spatial dimensions, the Lieb lattice possesses one flat band and two dispersive bands. 
Applying the periodical boundary conditions for tight binding Hamiltonian in Eq.~\eqref{Lieb tb Ham}, the Bloch Hamiltonian of Lieb lattice can be obtained:

\begin{equation}\label{Lieb Blach Ham}
    h(\mathbf{k})=2J
    \begin{pmatrix}
        0 & \cos(\frac{k_x}{2}) + i\delta\sin(\frac{k_x}{2}) & 0\\
       \cos(\frac{k_x}{2}) - i\delta\sin(\frac{k_x}{2}) & 0 & \cos(\frac{k_y}{2}) + i\delta\sin(\frac{k_y}{2})\\
        0 & \cos(\frac{k_y}{2}) - i\delta\sin(\frac{k_y}{2}) & 0\\
    \end{pmatrix}.
\end{equation}
Here we apply the notation $ a_{\mathbf{k}} = \cos(\frac{k_x}{2}) + i\delta\sin(\frac{k_x}{2}), ~b_{\mathbf{k}} = \cos(\frac{k_y}{2}) + i\delta\sin(\frac{k_y}{2}) $, and set lattice constant $a = 1$ for convenience. 
The band structure consists of two dispersive bands with dispersion $E_{\pm} =  \pm 2J \sqrt{1+\delta^2 + \frac{1-\delta^2}{2}\left(\cos k_x + \cos k_y\right)} $, and a complete flat band $E_0 = 0$ lying in between the dispersive bands. 

\begin{figure}[ht]
    \centering
    \includegraphics[scale=1]{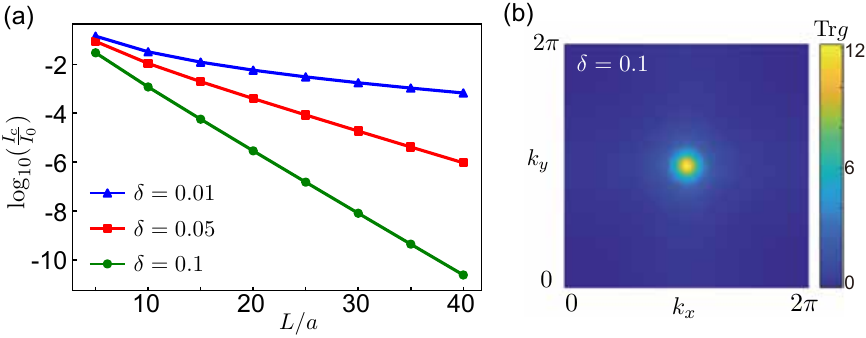}
    \caption{(a) Critical current $I_c/I_0$ in a 2D Josephson junction as a function of the junction length $L$, with $I_0=\frac{e\Delta_{\mathrm{sc}}}{\hbar}$. The decay length decreases as $\delta$ increases, which is qualitatively the same as the 1D result presented in the main text. 
    The parameters are $\mu_{\mathrm{s}}=\mu_{\mathrm{N}}=0$, $T_{\mathrm{l/r} \alpha}=100\Delta_{\mathrm{sc}}\delta_{\alpha \mathrm{A}}$, $k_\mathrm{B}T\!=\!0.1\Delta_{\mathrm{sc}}$, $t=100\Delta_{\mathrm{sc}}$, and $J =10^4\Delta_{\mathrm{sc}}$. The width of the weak link is $L_y = 200a$ and periodic boundary condition is assumed along $y$ direction.  (b) Distribution of the trace of quantum metric tensor $\mathrm{Tr} g$ in the first Brillouin zone for $\delta = 0.1$, and $J =10^4\Delta_{\mathrm{sc}}$.}
    \label{Ic2D}
\end{figure}


\section{Second Model for SC/FB/SC Josephson Junctions}
We have studied the Lieb lattices at both one- and two-spatial dimensions. In fact, the scenario that the interface states can mediate supercurrents with decay length controlled by quantum metric length can work for general flat band systems of quantum metric. We will provide a second model to illustrate the scenario. We consider a model of two spins and two valleys and the Hamiltonian with spin index reads 
\begin{equation}
    h_s(k) = -t_m [\lambda_x \sin(\alpha_k)+s\lambda_y\cos(\alpha_k)],
\end{equation}
with the time-reversal symmetry $h_{\uparrow}(k) = h_\downarrow^*(-k)$. 
Here $\alpha_k = \chi\cos(ka)$, $\lambda_i $ are the Pauli matrices in orbital basis, and $s=\pm1$ denote the spin $\uparrow$ and $\downarrow$. The Hamiltonian $h_s(k)$ contains two perfect flat bands with energy $\epsilon_k = \pm t_m$. The real parameter $\chi$ can tune the quantum metric. It is easily to the eigenstate at momentum $k$ 
\begin{equation}
    |u_+\rangle =\frac{1}{\sqrt{2}} [1,is e^{is\alpha_k}]^T,\quad |u_-\rangle = \frac{1}{\sqrt{2}} [-1, ise^{is\alpha_k}]^T.
\end{equation}
One can derive quantum metric $g(k)=\chi^2a^2\sin(ka)/4$. Therefore, we can alter the quantum metric by $\chi$ while keeping the band gap. 

\begin{figure}[t]
\centering
\includegraphics[scale=0.5]{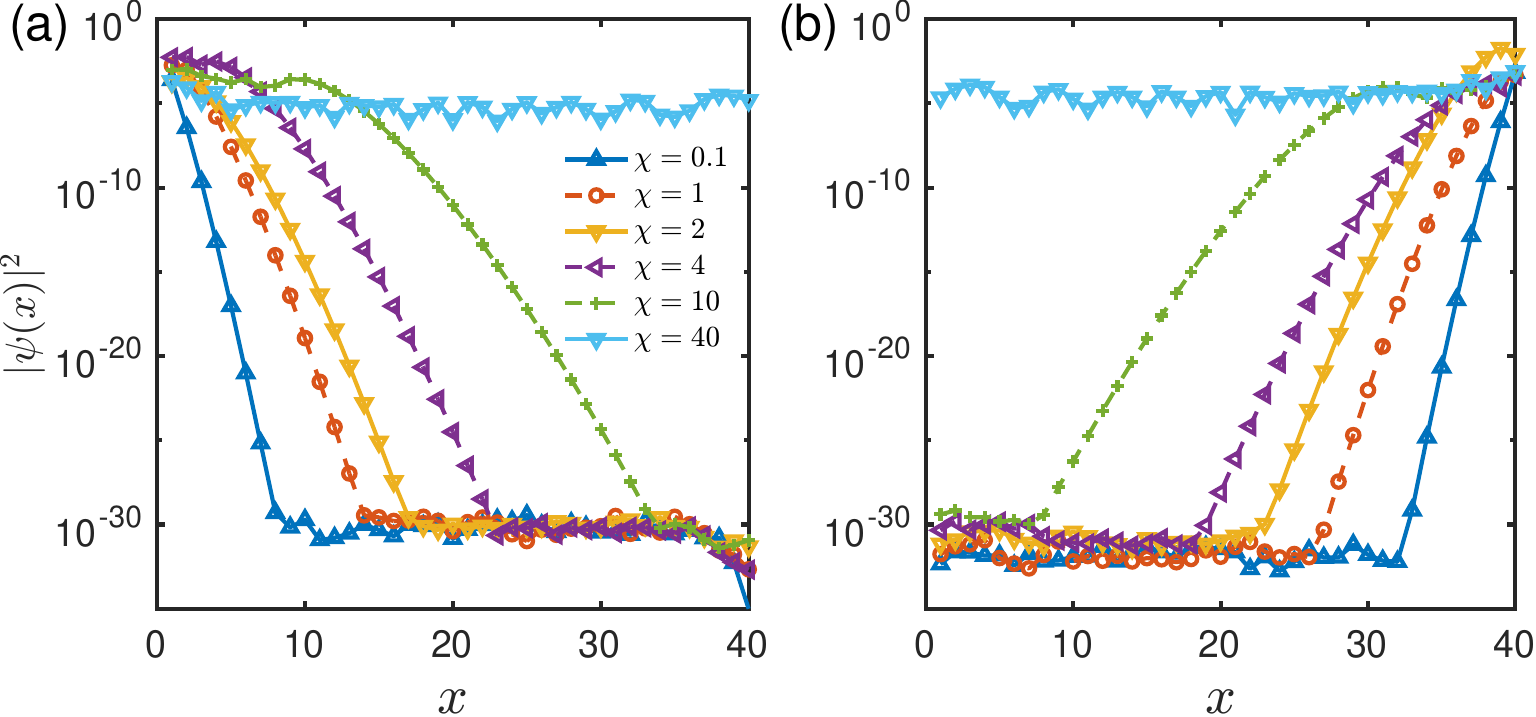}
\caption{The density profile of the density distributions of (a) left and (b) right interface bound states. The bound states decay exponentially into the weak links. When we increase $\chi$, the decay length becomes larger. Thus,
the bound states become more extensive when the quantum metric increases. We set the hopping integral $t$ of two superconducting leads as $t=2\Delta_\mathrm{sc}$ and the $\varphi_\mathrm{r}-\varphi_\mathrm{l} = \pi/2$. For the flat band model, we use $t_m=20\Delta_\mathrm{sc}$ and chemical potential $\mu_\mathrm{N}= 0$. In the contact Hamiltonian, we use $T=t_m =20\Delta_\mathrm{sc}$
}
\label{smfig:secondmodel}
\end{figure}

To construct the flat band junction, we need to convert the k-space Hamiltonian into the real space one. This can be done by the Fourier transformation. The spatial Hamiltonian involves long-range hopping due to the special form of the Hamiltonian. We can cut the spatial Hamitlonain to connect to two superconducting leads with the contract Hamiltonian as
\begin{equation}
    H_c = T \sum_{\alpha=\mathrm{A,B}} \sum_{\sigma } (c^\dagger_{\mathrm{l}\sigma}a_{\mathrm r\alpha \sigma} +c^\dagger_{\mathrm{r}\sigma}a_{\mathrm l\alpha \sigma} ),
\end{equation}
where $T$ is the coupling between the left/right superconducting leads and the flat band model. In Fig.~\ref{smfig:secondmodel}, we plot the profiles of left and right interface bound states. The larger $\chi$ is, the more extensive the bound states become. Thus, the decay lengths of the bound states are controlled by the quantum metric.

\maketitle

\end{document}